\documentclass[mathleft,fleqn,%
]{an2}
\usepackage{graphicx}
\usepackage{enumitem}
\usepackage[varg]{txfonts}
\usepackage{amssymb}

\usepackage[percent]{overpic}

\usepackage{natbib}
\bibpunct{(}{)}{;}{a}{}{,}
\newcommand{\deltanuF}{\mbox{$\Delta\nu$}}
\newcommand{\deltanu}{\mbox{$\langle \Delta\nu \rangle$}}
\newcommand{\numax}{\mbox{$\nu_{\rm max}$}}
\newcommand{\deltaP}{\mbox{$\Delta P$}}
\newcommand{\Kepler}{\textit{Kepler}}

\def\logg{$\log g$}
\begin{document}

\Pagespan{1}{}
\Yearpublication{2014}%
\Yearsubmission{2014}%
\Month{0}%
\Volume{999}%
\Issue{0}%
\DOI{asna.201400000}%

\title{PLATO \emph{as it is}: a legacy mission for Galactic archaeology}

\author{A. Miglio\inst{1,2}\fnmsep\thanks{Corresponding author:
        {a.miglio@bham.ac.uk}},   
 C. Chiappini \inst{ 3       },      
 B. Mosser \inst{ 4       },      
 G.\,R. Davies \inst{ 1 , 2     },      
 K. Freeman \inst{ 5       },      
 L. Girardi \inst{ 6       },      
 P. Jofr\'e \inst{ 7 , 8     },      
 D. Kawata \inst{ 9       },      
 B.\,M. Rendle \inst{ 1 , 2     },      
 M. Valentini \inst{ 3       },      
 L. Casagrande \inst{ 5       }
% ...
                }
\titlerunning{PLATO \emph{as it is}: a legacy mission for Galactic archaeology}
\authorrunning{A. Miglio et al.}

\institute{\it full list  of affiliations in appendix}

\received{XXXX}
\accepted{XXXX}
\publonline{XXXX}


\abstract{Deciphering the assembly history of the Milky Way is a formidable task, which becomes possible only if one can produce high-resolution chrono-chemo-kinematical maps of the Galaxy. 
Data from large-scale astrometric and spectroscopic surveys will soon provide us with a well-defined view of the current chemo-kinematical structure of the Milky Way, but will only enable a blurred view on the temporal sequence that led to the present-day Galaxy.
As demonstrated by the (ongoing) exploitation of data from the pioneering photometric missions CoRoT, {\it Kepler}, and K2, asteroseismology  provides the way forward: solar-like oscillating giants are excellent evolutionary clocks thanks to the availability of seismic constraints on their mass and to the tight age-initial-mass relation they adhere to. 
In this paper we identify five key outstanding questions relating to the formation and evolution of the Milky Way that will need precise and accurate ages for large samples of stars to be addressed, and we identify the requirements in terms of number of targets and  the precision on the stellar properties that are needed to tackle such questions.
By quantifying the asteroseismic yields expected from PLATO for red-giant stars, we demonstrate that these requirements are  within the capabilities of the current instrument design, provided that observations are sufficiently long to identify the evolutionary state and allow robust and precise determination of acoustic-mode frequencies.  This will allow us to harvest data of sufficient quality to reach a 10\% precision in age. This is a fundamental pre-requisite to then reach the more ambitious goal of a similar level of accuracy, which will only be possible if we have to hand a careful appraisal of systematic uncertainties on age deriving from our limited understanding of stellar physics, a goal which conveniently falls within the main aims of PLATO's core science.
We therefore strongly endorse PLATO's current design and proposed observational strategy, and conclude that PLATO, {\it as it is}, will be a legacy mission for Galactic archaeology.}

\maketitle
\section{What this paper provides}
This paper spells out outstanding questions in Galactic astronomy that will still be unresolved in 10 years' time; and explains in detail how the ESA PLATO Mission\footnote{\texttt{http://sci.esa.int/plato/}} \citep{Rauer2014}, in its current form (design specification\footnote{Satellite with 24 cameras and a nominal 4-year observing run,  built and verified for an in-orbit lifetime of 6.5 years, as described in the \emph{PLATO Definition Study Report.}}) will be able to address these challenges. 

We specify in detail the requirements on numbers of targets, estimated stellar properties (including precise ages), as well as the pointing strategy requirements needed to fullfil the Galactic archaeology goals. 
The breakdown of this paper is as follows:
\begin{itemize}
\item An introduction to Galactic archaeology is given in Section \ref{sec:intro}, while key limitations and outstanding questions in the field are identified in Section \ref{sec:openquestions}.
\item The need for high-precision stellar ages and the role of asteroseismology is reviewed in Section \ref{sec:whysismo}, and the requirements on the performance of PLATO as a Galactic archaeology mission are listed in Section \ref{sec:desiderata}.
\item The expected asteroseismic yields for  PLATO (red-giant stars) are discussed in Section \ref{sec:seismo}, and the impact of the duration of the observational campaigns on the number of stars with detectable oscillations, and on the precision of the inferred stellar properties (in particular age) is reported in Section \ref{sec:star_prop}.  
\item  Additional constraints on stars that allow synergies with PLATO's asteroseismic data, such as distances, extinction maps, and surface gravities (hence synergies with spectroscopic surveys) are presented in Sections \ref{sec:distance} and \ref{sec:spectroscopy}.  
\item Finally, a brief summary is given is Section \ref{sec:summary}.
\end{itemize}

\section{Introduction}
\label{sec:intro}
Galaxies are complex systems, with dynamical and chemical substructures, where several competing processes such as mergers, internal secular evolution, gas accretion and gas flows take place. 
Galactic archaeology of the Milky Way aims at taking advantage of the fact that for our Galaxy all these processes can potentially be disentangled thanks to the use of high dimensionality maps obtained by combining kinematic, chemical, and age information for stars belonging to the Galactic components and substructures 
\citep[e.g.][]{Matteucci2001,Freeman2002,Pagel2009,Rix2013}. That researchers on Galactic science are convinced this is the way forward has become clear by the large investments in missions such as Gaia \citep{Gaia_Prusti2016}, as well as comparatively large efforts devoted to large-scale ground-based photometric and spectroscopic surveys \citep{Turon2008}.  
 
Deciphering the assembly history of our Galaxy now seems a reachable goal. The complexity of the data already in hand (for instance from combining current spectroscopic information with the Gaia-TGAS sample; see e.g. \citealt{Michalik2014}), makes it clear that only for our Galaxy will one be able to achieve this goal in the foreseeable future. However, it has also become evident to Galactic archaeologists that one of the main pieces of the puzzle is still missing: \emph{precise ages for stars, covering large volumes of the Milky Way} \citep[e.g., see][and references therein]{Freeman2012, Chiappini2015b}. The latter requirement implies the use of red giants as tracers because these are bright enough to be observed at large distances, thus offering the opportunity to truly map the Galaxy. 

The ESA Gaia satellite will soon deliver a 6-D map\footnote{The star's position plus 3-dimensional velocities. These are complemented by further dimensions in chemical space.} of $10^5$ stars and a 5-D map\footnote{Position plus tangential velocity.} of more than one billion stars throughout our Galaxy \citep{Gaia_Brown2016, Cacciari2016}. Additional crucial information, both on velocities and chemical abundances, will come from several ongoing/planned spectroscopic surveys such as RAVE \citep{Steinmetz2006,Kunder2017}, SEGUE-2 \citep{Yanny2009,Eisenstein2011}, APOGEE \citep{Majewski2015,Majewski2016}, Gaia-ESO \citep{Gilmore2012}, LAMOST \citep{Cui2012}, GALAH \citep{deSilva2015,Martell2017}, WEAVE \citep{Dalton2014},  4MOST \citep{deJong2014}, DESI \citep{DESI2016a, DESI2016b} and MOONS \citep{Cirasuolo2014}. However, astrometric and spectroscopic constraints alone will not enable a precise and accurate estimate of red-giant ages\footnote{Age-dating of field red giants from isochrone fitting to observations in an HR diagram is known to be a challenge as small uncertainties on the observational constraints lead to large uncertainties on the mass (and hence age) estimates. Other recent, and more indirect methods using surface abundances of carbon and nitrogen \citep[e.g][]{Martig2016} are not able to deliver ages of the precision aimed for here \cite[e.g.][]{Salaris2015,Lagarde2017}, while spectroscopic data-driven approaches \citep{Ness2016,Casey2017} do require high-precision training sets to be able to deliver precise ages.}: here is where PLATO will play a fundamental and unique role. With PLATO it will finally be possible to have large samples of red giants, thus cover a large volume of the Galaxy, for which precise ages will be known.

\subsection{Scientific motivation}
\label{sec:openquestions}

The knowledge of age for distant stars is key to helping to disentangle the multi-dimensional problem of Galaxy assembly. Some of the pressing questions related to the origin of the oldest Galactic components such as the halo, the thick disk, and the bulge do require an age map of the oldest stars towards several directions in the Galaxy. Breakthroughs are expected if ages are known to the 10\% precision level, especially at old ages (i.e. covering the first 2-4 Gyrs of the evolution of our Galaxy).  Moreover, ages with a 10\% precision for stars in the Galaxy will let us accurately interpret the evolution of the Milky Way in the context of the evolution of disk galaxies observed at high redshift.

Indeed, the important formation phase in high-$z$ disk galaxies appears to have been between about 12 and 8 Gyr ago: after that time,
thin disk formation appears to continue relatively sedately to the present.  In this early 
interval of about 4 Gyr, the basic structure of bulges/halo, thick and thin disks in disk galaxies 
as we see them now was established, as suggested by many theoretical models \citep[e.g.][]{Jones1983,Steinmetz1994,Noguchi1998,Abadi2003,Sommer-Larsen2003,Brook2004,Bournaud2009,Gibson2009,Bird2013,Guedes2013,Kawata2016}. This seems to be also the case in the Milky Way \citep{Chiappini1997,Chiappini2009,Minchev2013,Minchev2014,Kubryk2015,Snaith2015} where current data suggest the 
 thick disk formation started at $z \sim$ 3.5 (12 Gyr ago) while the thin disk began to form at $z \sim$ 1.5 (8 Gyr ago) \citep[e.g.][]{Fuhrmann2011,Haywood2013,Bensby2014,Bergemann2014,Robin2014}.  
The modern aim in Galactic archaeology is to build an extensive chemo-kinematical-age map of the Galaxy, and finally tackle the still open questions in the field. Some of these are:

\begin{enumerate}

\item What is the origin of the two chemically different populations of the Galactic disk, i.e. the $\alpha$-rich\footnote{The $\alpha$ elements are named as such because their nuclei are multiples of $^4$He nuclei ($\alpha$ particles).} and $\alpha$-poor disks?\footnote{These terms are often used in the literature to refer to the [$\alpha$/Fe] ratio, where [X/H] = log(X/H) - log(X/H)$_{\odot}$. An $\alpha$-rich population is made of stars that have [$\alpha$/Fe]$>$0.1-0.2, depending on metallicity. This in turn is indicative of a population mainly enriched by core-collapse supernovae, and hence formed on short timescales. } The current observational evidence suggests the $\alpha$-rich disk to be systematically older than the $\alpha$-poor disk component \citep{Fuhrmann2011,Martig2016,Ness2016}. Is there a smooth transition from an $\alpha$-rich to an $\alpha$-poor disk \citep{Bovy2012}? Or is there a discontinuity (for instance caused by a drop in the star formation rate) which would imply the thick and thin disk are two genuine discrete Galactic components with different chemical evolution histories \citep{Chiappini1997,Reddy2006}? 

\item What are the age-velocity and age-metallicity relations in the whole disk, bulge, and halo? Even for the local volume, both relations are still a matter of debate \cite[e.g.][]{Quillen2001,Holmberg2007}.
The radial and vertical variations of these two relations are reflected in the disk chemical abundance gradients \citep[e.g.][]{Cheng2012,Boeche2013,Boeche2014,Anders2014,Anders2017b,Hayden2014,Mikolaitis2014,Jacobson2016}, as well as on variations of metallicities and abundance ratios with Galactocentric distance and Galactic height \citep{Hayden2015,Rojas-Arriagada2016,Anders2017b}. All these constitute key constraints to scenarios of disk, bulge, and halo formation. Which of these Galactic components have formed inside-out, and which have formed outside-in?  

\item When was the bar formed? How did the bar grow? Has the $\alpha$-poor disk  shrunk vertically with time or were older stars heated up by interacting with the bar, spiral arms and/or giant molecular clouds? A map of the evolution of stellar velocity dispersions in the disk would provide important answers to questions related to the origin of the thick disk and on the main sources of heating in the disk (mergers, molecular clouds, radial migration) in chronological order. Current evolutionary models \citep[e.g.][]{DiMatteo2013,Grand2016,Athanassoula2017} are in desperate need for these tighter constraints. 

\item Does the bulge just come from the instability of the inner thin and thick disk components, or is there a
significant classical merger-generated bulge \citep[see][for recent reviews]{Bournaud2016,Shen2016,Nataf2016,Naab2016}?
How is the formation of the thick disk connected to that of the bulge? Are these multi-populations responsible for the multi-peak metallicity distribution unveiled by modern data of the bulge regions \citep[e.g.][]{Babusiaux2016}?  What is the contribution of the inner disk to the bulge/bar \citep[e.g.][]{DiMatteo2014}? What is the age distribution of the multi-peak metallicity distribution components observed in the inner regions of the Galaxy \citep[e.g.][]{Bensby2017}? 

\item How important is radial migration? Is it so intense that it would be able to partially delete the Galactic archaeology fossil records?  
What is the nature and the role of the spiral arms and bar as sources of radial migration? Is migration caused by transient \citep{Sellwood2002} or long-lived \citep{Minchev2010} patterns?
How much of the radial migration is also caused by mergers \citep{Quillen2009,Bird2013}?  In relation to the disk and its merger history, 
how have the abundance gradients today observed in the thin and thick disks evolved? Were these gradients significantly affected by radial migration? Was the flaring of the thin disk stronger in the past \citep{Amores2017}? As recently illustrated by \citet{Minchev2017} and references therein, ages for large samples of stars are needed to be able to tackle the above questions. 

\end{enumerate}

Researchers in the Galactic archaeology field are now convinced that combining asteroseismic, astrometric,  and spectroscopic observational constraints provides the way forward in the field \citep[see e.g.][for a recent overview]{Noels2016}. Modern data will be rich in details and hence complex. The ultimate challenge will be that of building models able to interpret this rich dataset, and finally shed light on all the above questions.

\subsection{Why is asteroseismology needed?}
\label{sec:whysismo}
One of the main challenges of Galactic archaeology in the PLATO era is to reveal the Galaxy assembly and evolution history via the age, chemical composition, and kinematics of stars in a large fraction of the volume in the Milky Way. 
Chemical properties and radial velocities can 
already be measured (at different levels of precision) by surveys such as SEGUE, RAVE, Gaia-ESO, 
APOGEE, LAMOST, GALAH and, in the near future, 
WEAVE, 4MOST and MOONS. The radial velocity and chemical properties for  bright stars and transverse kinematics for all the stars detected by Gaia will soon be available from the upcoming Gaia data releases. 
These large datasets will ensure we will have by $\sim$2025 a good picture of the \emph{current} chemodynamical structure of the Milky Way. However, the
critical chronological information that we need for Galactic archaeology to understand the formation
and evolution of the Milky Way will still be missing.

\begin{figure}
\includegraphics[width=\linewidth]{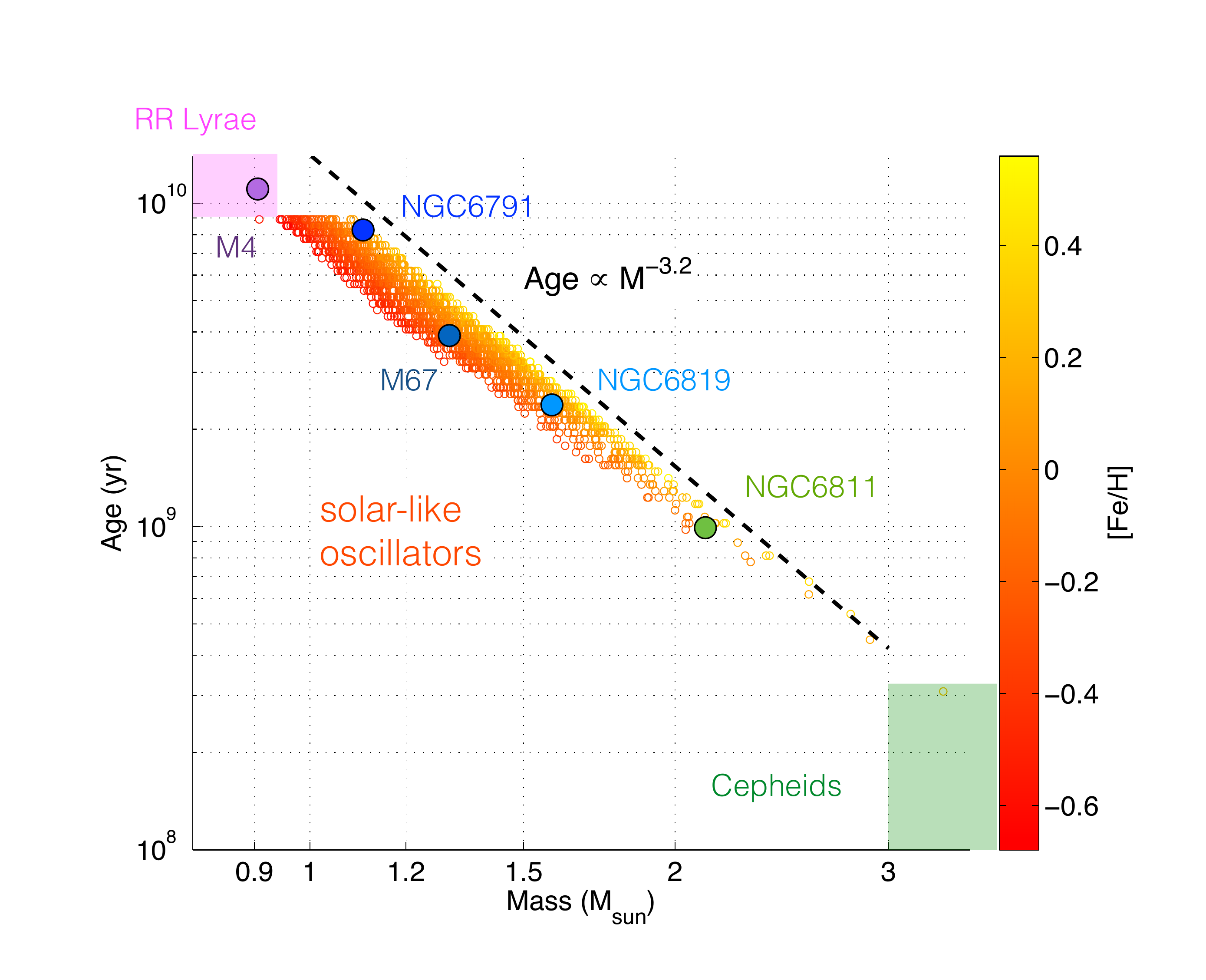}
\caption{Age-mass-metallicity relation for red giants in a {\sc trilegal} \citep{Girardi2005} synthetic population representative of thin-disk red-giant-branch (RGB) stars observed by \emph{Kepler}. The dashed line indicates the average power-law relation between age and mass of RGB stars. Given their extended mass range and the tight age-mass relations, solar-like oscillating giants (dots) probe the full history of the Milky Way.  The asteroseismic age scale is currently being validated primarily thanks to the detection of oscillations in giants belonging to open and globular clusters observed by {\it Kepler} and K2 \citep{Brogaard2012, Miglio2016, Handberg2017, Stello2016, Brogaard2016, Sandquist2016, Molenda2014, Arentoft2017}. Classical pulsators in similar evolutionary phases (Cepheids and RR Lyrae stars) are also indicated in the diagram.}
\label{fig:agescale}
\end{figure}

Asteroseismology, i.e. the study and interpretation of, and the astrophysical inference from global oscillation modes in stars, provides the way forward. 
Along with enabling exquisite tests of stellar models,  pulsation frequencies of the solar-like oscillators may be used to place tight constraints on the fundamental stellar properties, including radius, mass and evolutionary state (see, e.g., \citealt{Chaplin2013}, \citealt{Christensen-Dalsgaard2016}, \citealt{Hekker2016}, and references therein). 
Stellar mass is a particularly valuable constraint in the case of giants, since for these stars there is a very tight relation between age and mass. 
The age of low-mass red-giant stars is largely determined by the time spent on the main sequence, hence by the initial mass of the red giant's progenitor ($\tau_{\rm MS} \propto M/L(M) \propto M^{-(\gamma-1)}$, with $\gamma  \sim 4$, where $L$ is the typical luminosity of the star on the main sequence, e.g. see \citealt{Kippenhahn2012}). With asteroseismic constraints on the stellar mass, it is now possible to infer the age of thousands of  individual stars, spanning the entire evolution of the Milky Way (see Fig. \ref{fig:agescale}). 

One of the most convincing (and highly-regarded) statements about the importance of asteroseismology for Galactic archaeology can be found in the  ESO-ESA Working groups Report 4  on  Galactic populations, Chemistry and Dynamics \citep{Turon2008}. This working group was requested by ESO and ESA to consider projects that would complement the Gaia mission. One of the recommendations made to ESA was: ``Asteroseismology: this is a major tool to complement Gaia with respect to age determinations. ESA should encourage the community to prepare for a next-generation mission, which would sample the different populations of the Galaxy much more widely than CNES-ESA's CoRoT and NASA's {\it Kepler}'': \emph{PLATO is the mission that can deliver long-sought constraints to models of the Milky Way assembly and evolution.}

The combination of Gaia and spectroscopic surveys will be able to tell us the difference between photometrically defined thick and thin disks vs. chemically defined $\alpha$-rich and $\alpha$-poor disks \citep[for a discussion regarding the various definitions of the thick and thin disks see e.g.][]{Minchev2015,Kawata2016}. 
 Age information of turn-off stars will be available in the Gaia era. However, these stars are intrinsically faint, preventing a large volume coverage of the Galaxy \citep[e.g. see][]{Cacciari2016}. For giants the current age estimates are very uncertain (for instance those based on C and N spectral features,  e.g. \citealt{Masseron2015, Martig2016}) and more precise age estimates mainly rely on relatively small asteroseismic data sets from {\it Kepler} \citep{Borucki2010}, K2 \citep{Howell2014} and CoRoT \citep{Baglin2006, Corot2016}. What is needed is more reliable and homogeneously derived age information for a much larger number of stars, covering larger volumes of the Milky Way.

It has now been demonstrated that precise and more accurate (although still stellar-model dependent) ages can be inferred for the solar-like pulsating red giants observed by the space-borne telescopes CoRoT, {\it Kepler}, and K2 (see e.g. \citealt{Miglio2013, Casagrande2016, Anders2017a, Rodrigues2017}). The combination of chemical compositions from spectroscopic surveys with distances and motions from Gaia and ages from asteroseismic data, on large samples of stars, will allow us to comprehensively study chemodynamical distributions and their time evolution in different directions of the Milky Way.

A recent application demonstrating the potential of such a combination was recently presented by \citet{Anders2017b}, where around 400 stars from just two of the CoRoT fields that have measurements with APOGEE spectra (and hence velocity and chemical information) have been used to estimate the evolution of the abundance gradients in the thin disk in the last 6-8 Gyrs, a long-sought constraint to the chemical evolution of the Milky Way. 
A further example is given by the discovery of the so-called young-$\alpha$-rich stars \citep{Chiappini2015,Martig2015}, i.e. stars with masses implying young ages, but which feature an overabundance in $\alpha$-elements, typical of old stars. It is still unclear whether the large numbers of young-$\alpha$-rich stars found so far is compatible with the assumption of them being just blue stragglers, rather than genuine young stars \citep{Jofre2016,Fuhrmann2017}. In addition, it will finally be possible to map the thick and thin disk components also with respect to their age, which can turn out to be key, as the overlap in metallicities and kinematics blur our understanding of the two components. The precise measurement of the existence or not of an age gradient in the thick disk can also put strong constraints to its assembly \cite[e.g.][]{Minchev2015}.

All of these crucial constraints will allow us to quantify the importance of stellar radial migration in the formation of the Milky Way, otherwise difficult to quantify from first principles. This will represent invaluable information not only for the formation of the Milky Way, but also for the formation of spiral galaxies in general.

\subsection{What can PLATO do for Galactic archaeology that previous missions could not?}

While pioneering photometric space missions such as CoRoT, {\it Kepler} and K2 have demonstrated the enormous potential of seismology for stellar populations studies, they all have limitations relating to spatial and temporal coverage. {\it Kepler} provided a unique survey in a 105 deg$^2$ area, continuously observed during  four years. This survey, however, provides a limited census of the Milky Way's properties. 

The K2 and TESS\footnote{See e.g. \citet{Campante2016} for predictions of the asteroseismic yields of TESS.} \citep{Ricker2015} missions provide or will provide, respectively, a large-area and a whole-sky survey. Their results for studying the Milky Way's properties are limited by a 
short observation duration; the resulting frequency resolution limits the seismic analysis of evolved stars in numerous cases compared to what can be achieved by PLATO (see Section \ref{sec:seismo}). 
The results provided by CoRoT were based on a  good compromise between the extent of the survey and the observation duration, but were limited by the photon-noise resulting from its 28-cm diameter mirror, and limited sky coverage.  

\emph{PLATO is the only planned mission that can overcome these limitations}, and therefore will have an enormous impact in the field of Galactic archaeology in several ways, namely:
\begin{enumerate}
\item It will provide constraints on the properties of large ensembles of stars (in the giant phase, but crucially also on the main-sequence
and subgiant phase) enabling stringent tests of stellar structure and evolution models, leading to an improved accuracy on predicted stellar parameters and yields,
\item It will explore connections between populations of exoplanets and those of the host stars, 
\item It will allow to address important open questions in Galactic archaeology and will deliver the first chrono-chemo-kinematical map of the Milky Way.
\end{enumerate}

In the following section we outline the specific asteroseismic performance requirements (e.g. number of stars, their spatial distribution, precision on age) needed to address the outstanding questions in Galactic archaeology. We then explore in detail (Section \ref{sec:seismo}) what PLATO is expected to achieve in terms of seismic yields for red-giant stars, including estimates on the precision on inferred stellar properties depending on the duration of the observational campaigns.

\section{Performance requirements for a PLATO-Galactic archaeology mission}
\label{sec:desiderata}
The distance range to be covered by oscillating red-giant stars and the need for precise ages for these objects set the basic requirements on the limiting magnitude, the duration of observations, and the level of seismic analysis (both data analysis and modelling) required for Galactic archaeology in the PLATO era. 

\begin{figure}
\includegraphics[width=\linewidth]{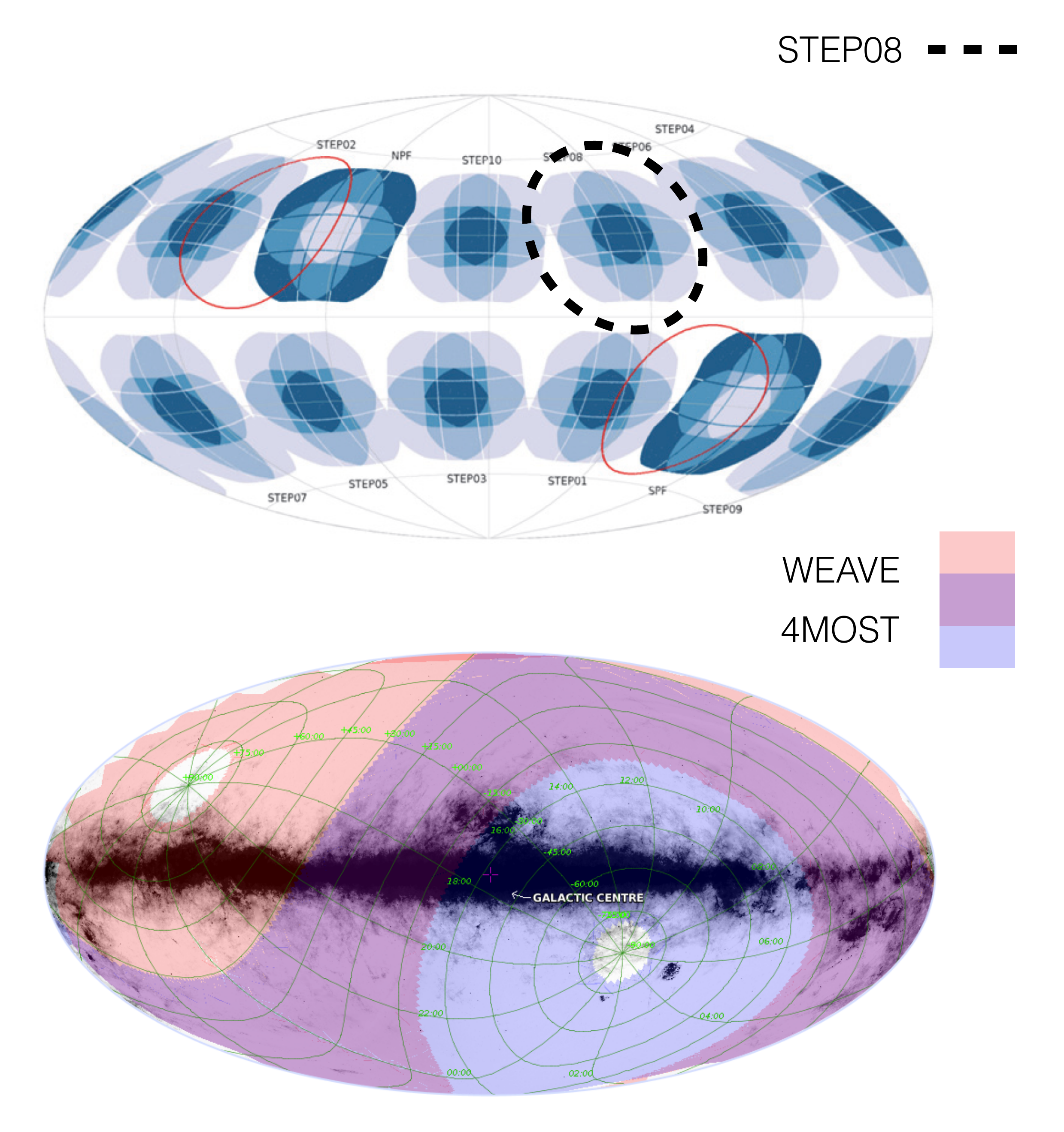}
\caption{{\it Upper panel:} Projection of the two preliminary long-duration (LD) fields (Southern Plato Field, Northern Plato Field) and ten step-and-stare fields (STEP01 to STEP10), all centred at $|b| = 30$, in the Galactic reference frame. The red line is the LD pointing requirement limit. The LD fields are colour-coded on an inverted scale. In the current instrument design various parts of each field are monitored by 24, 18, 12 or 6 cameras (as indicated by different colours). The field selected for this study (STEP08) is encircled by a thick dashed line (Figure taken and adapted from the PLATO Definition Study Report).
{\it Lower panel:} expected sky coverage of the forthcoming spectroscopic surveys 4MOST and WEAVE superposed on an IRAS map of the sky \citep{IRAS2002}.}
\label{fig:fields}
\end{figure}

\begin{figure*}
\includegraphics[angle=-90, width=\linewidth]{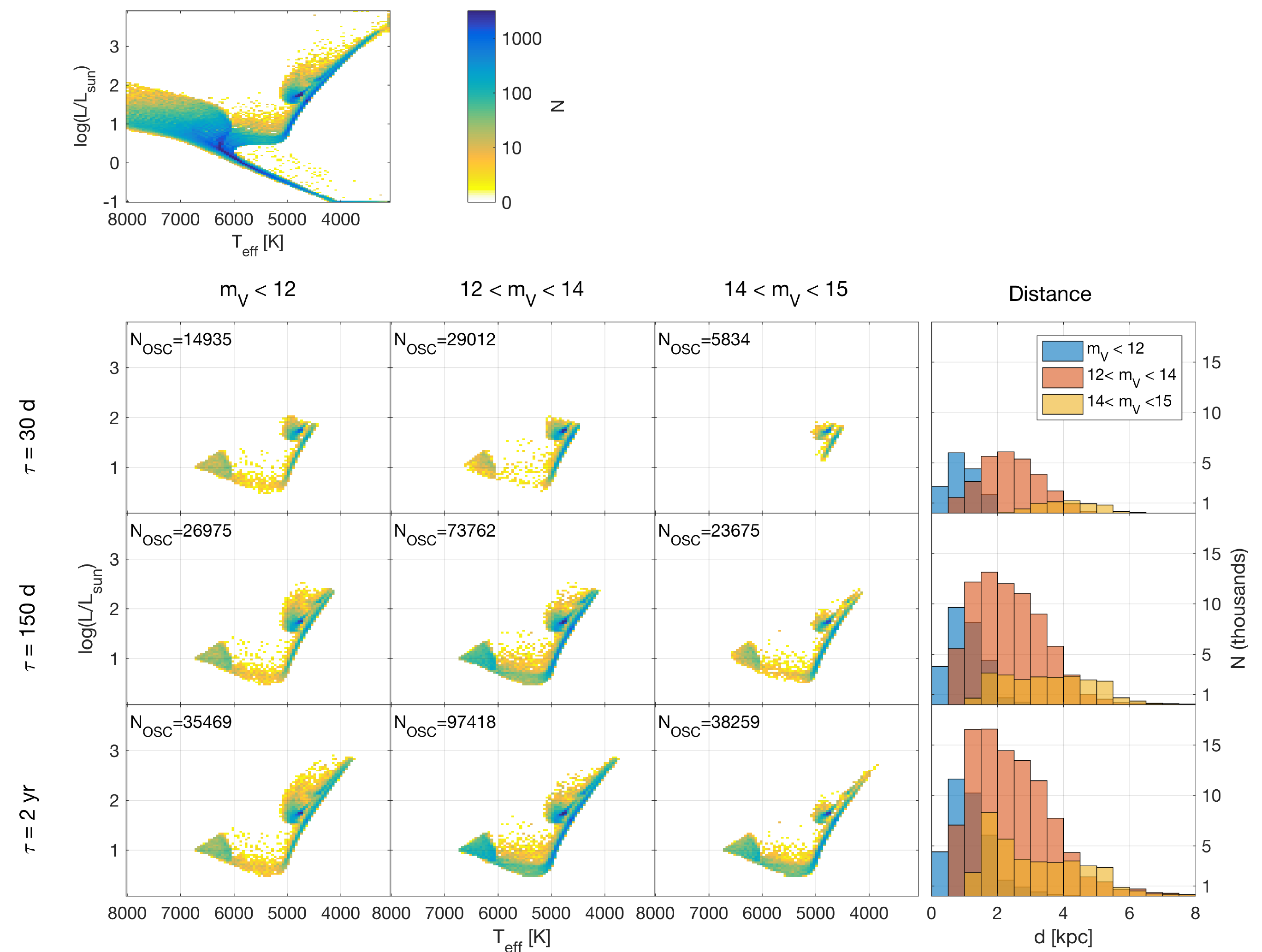} 
\caption{{\it Upper panel:} HR diagram of the synthetic population simulated with {\sc trilegal} in the PLATO field STEP08. Colour represents the number of stars per $T_{\rm eff}$-$\log{L}$ bin. {\it Lower panel:} In each row we show the HR diagram of stars with detectable oscillations and in different magnitude bins, with N$_{\rm OSC}$ indicating the approximate number of stars with detectable solar-like oscillations. Our predictions are limited to stars with oscillation frequencies lower than $\sim$ 800 $\mu$Hz, hence primarily to stars in the red-giant phase of evolution; see the main text for details. The distance distribution of such stars is presented in the right-most panel. Different rows illustrate the effect of increasing the duration of the observing run, $\tau=$ 30 d, 150 d and 2 yr (first, second, and third row, respectively).}
\label{fig:cmd}
\end{figure*}

To ensure that the PLATO mission exploits its full legacy value also for the field of Galactic archaeology, two main requirements need to be met: a) the observing runs need to be long enough to provide age uncertainties below $\sim$ 10\% at the oldest ages (see Section \ref{sec:seismo}), and b) a strategic field placement is needed, enabling mapping of both the azimuthal and vertical structures of the Galactic components. The \emph{current} PLATO proposal of long and short runs, as well as the planned field placement (see Fig. \ref{fig:fields}), fulfils these two Galactic archaeology requirements for the following reasons:
\begin{itemize}

\item[$\bullet$]  {\it Radial, and vertical variations of chemo-kinematic properties of the thick and thin disks:} From current spectroscopic survey data,  
we know already that the properties of the (chemically defined) 
Galactic thin and thick disk change with radius and height. These changes are critical indicators of how the
thin and thick disks were assembled at high redshift and subsequently evolved. To cover a 
useful range in radius, we need to study stars out to at least 5 kpc from the Sun.  
For red clump giants, and negligible extinction, this corresponds to magnitudes of about $m_{\rm V} = 14$. Results discussed in Section \ref{sec:simulations} and Fig. \ref{fig:cmd} show that this criterion is easily met and surpassed given the current mission design.

\item[$\bullet$]  {\it Radial and azimuthal variations of chemo-kinematic properties of bulge and inner disk:} Given that PLATO will be able to detect oscillations in red-giant stars down to magnitudes of at least $m_{\rm V} \sim$15, as shown by our simulations in Section \ref{sec:simulations} and  Fig. \ref{fig:cmd}, one should consider fields within the Galactic bulge/bar in order to establish more
accurately the bulge history and its relation to the inner disk. Furthermore, given the radial and likely azimuthal dependence of chemo-kinematic properties owing to the presence of the bar and spiral arms, 
it is highly desirable to acquire data for giants in several Galactic fields\footnote{The expected PLATO field-of-view at each pointing is 2232 deg$^2$.} covering different Galactic longitudinal directions.  It will be valuable to have, for instance, a) two inner fields near Galactic longitude $l=\pm 20$ deg and $|b|$ = 30 deg, respectively, thus sampling the inner-disk and bulge regions, and b) another two fields at $l = 90$ deg or 270 deg and $l=180$  deg ($|b|$ $\sim$30 deg) to well sample the whole disk. Because the field diameter is $\sim 45$-deg wide, at $|b|$ = 30 deg one will still reach objects close to the non-heavily extinct Galactic plane (sampling $|b|$ down to 10-15 degrees). By adding extra fields covering even lower latitudes ($b =\pm$ 4 deg) one would be able to better explore the Bulge structure (the long bar at $l = +$15-20 deg at $b=$4-5 deg -- \cite{Wegg2015}, as well as the Baade's window at $b=-4$ deg).

\item[$\bullet$] {\it Mono-age populations:} The fast evolution anticipated for the earliest phases of our Galaxy (building the halo, bulge, thick disk and inner-thin disk early on, around 1 to 4 Gyr after the Big Bang) defines the accuracy of the ages 
that would be desirable for studying Galactic archaeology in this early epoch. 
An age precision of  about 10\% is required to 
follow in detail the formation and early evolution of the thin and thick disks 
of our Galaxy, and in particular to identify the transition between $\alpha$-rich and $\alpha$-poor disks over large Galactic volumes
(ideally  $0<R_{\rm gal}<20$ kpc, and $0<|z|<3$ kpc). This requirement is met and surpassed for a duration of the observations of the order of 5 months or more, as will be shown in the next Sections.

\item[$\bullet$] {\it The age-velocity dispersion relation:} In addition, with accurate age information (with uncertainties below $\sim$1 Gyr for the oldest age bins) for $\alpha$-rich and $\alpha$-poor stars, and with a large volume coverage of the disk 
($3<R_{\rm gal}<12$ kpc and $0<|z|< 3$ kpc), 
it will be possible to measure  the radial scale-length and vertical scale height as a function of Galactocentric radius for mono-age disk populations. The current suggested fields, centered on $b=$30 but reaching $b \sim5$ deg, are ideal for this. In the redshift interval between $z = 3$ ($\sim$13 Gyr) and $z = 1$ ($\sim$8 Gyr), the velocity dispersion of the gas in star-forming disk galaxies decays from about 80 km~s$^{-1}$  to about 30 km~s$^{-1}$  
(e.g. \citealt{Wisnioski2015}). Maps of the age vs. velocity dispersion at the different locations of the Galactic disk would enable the detection of a  sudden change of the radial velocity dispersion at the oldest ages, in case the same happens for our Galaxy. 

\end{itemize}

With the above requirements fulfilled, PLATO will represent a legacy for Galactic archaeology, uncovering the Milky Way assembly history, which no other mission is able to accomplish in the foreseable future. These data will enable the construction of maps of the radial and vertical metallicity gradients and of the width and skewness of the metallicity distribution function at different locations, for mono-age populations of stars. This will provide strong constraints on the relevance of radial migration, which is closely related to the nature and strength of the spiral arms and bar, to the birth place of the Sun as well as to the merger history of the Galaxy. By comparing these data with advanced chemodynamical simulations, it will be possible to re-construct the metallicity distributions of mono-age populations and quantify the impact of radial migration along the Milky Way evolution. The inferred metallicity distribution of star-forming regions at different epochs will be compared with the metallicity distribution of high-redshift galaxies which will soon be more accurately observed with Adaptive Optics and Integral Field Unit data with 30-m-class telescopes \citep[e.g. current state-of-the-art with KMOS/VLT seen in][]{Wuyts2016}.

As {the PLATO input catalogue} will be based on Gaia data, one will have all the information needed for modelling the selection biases involved. In addition, possible biases related to the detectability of solar-like oscillations can be accounted for \citep[e.g., see][]{Chaplin2011}.

\begin{figure*}
\centering
\includegraphics[width=.75\linewidth]{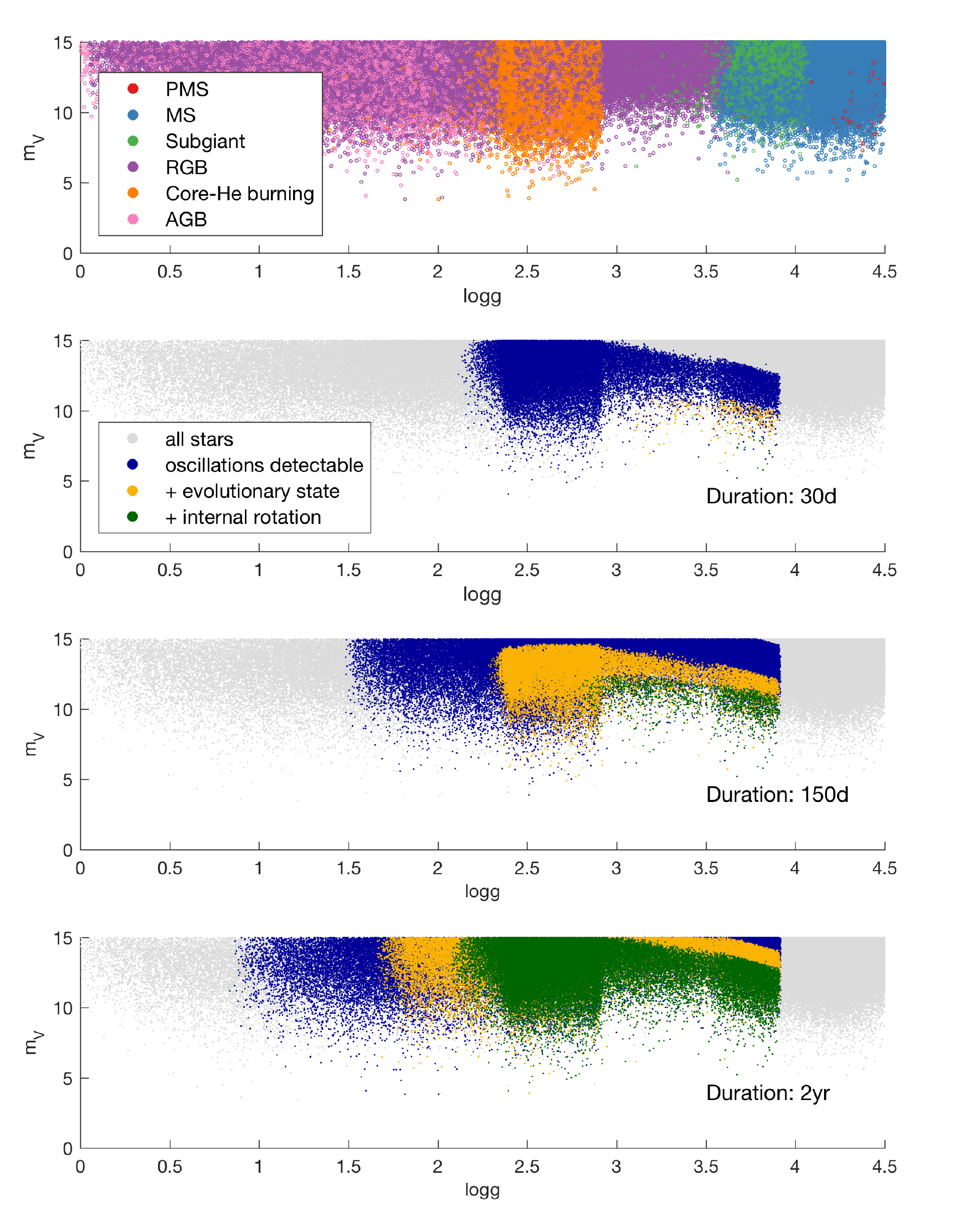}
\caption{Stars in the synthetic populations (upper panel of Fig. \ref{fig:cmd}) are presented in a surface gravity versus apparent $V$-band magnitude plot. The uppermost panel illustrates the location of stars in different evolutionary states (as defined in \citealt{Bressan2012}), from pre-main-sequence (PMS) to asymptotic-giant-branch (AGB) stars. The other three panels show the expected seismic yields as a function of the duration of the observations, from top to  bottom: yields for observations with durations of 30 d, 150 d and 2 yr. Each star in the population is coloured according to the seismic information that can be extracted: gray: no detections, blue: oscillations are detectable (hence \deltanu\ and \numax can be measured), yellow: evolutionary state, based on the detection of the gravity-mode period spacing, can also be inferred, and green: rotationally-split pulsation modes can be measured, hence information on the internal rotational profile can also be inferred.  Our predictions are limited to stars with oscillation frequencies lower than $\sim$ 800 $\mu$Hz, hence primarily to stars in their red-giant phase of evolution (which are not part of PLATO's core-target list, see the main text for details).}
\label{fig:yields}

\end{figure*}

\section{Expected seismic performance}
\label{sec:seismo}

We make use of the experience acquired with the analysis of  \Kepler\ observations to quantify the expected performance for PLATO. We focus on evolved stars, which represent ideal probes of Galactic structure, primarily thanks to their intrinsic brightness (see Section \ref{sec:intro}), and whose oscillations have low-enough frequencies to be detectable using PLATO long-cadence data. We refer to \citet{Rauer2014} for a discussion about the seismic performance expected for {solar-like pulsating} main-sequence stars.

\subsection{Simulating PLATO fields}
The proposed PLATO fields spam 48.5-deg wide squares on the sky. We simulate one of these fields, STEP08 centred at $(l, b)=(315\,  \mathrm{deg}, +30\,  \mathrm{deg})$ (see Figure \ref{fig:fields}), using the {\sc trilegal} tool \citep{Girardi2005, Girardi2012}. The entire field is initially split into small (0.8 deg$^2$) subareas by means of the healpix \citep{healpix} method. For each subarea, the mean extinction and its dispersion are computed from \citet{SFD} extinction maps, and later distributed along the line-of-sight as if the extinction were caused by a diffuse exponential dust layer with a vertical scale height of 110~pc. In this way, nearby dwarfs are little affected by extinction, while the distant giants in practice have the same distribution of extinction values as provided by  \citet{SFD}. The {\sc trilegal} model contains stars in the thin and thick disks, and halo, drawn from extended grids of stellar evolutionary and atmosphere models. They follow reasonable star formation histories and age-metallicity relations, and density distributions with well-accepted functional forms but with their total densities re-scaled so that the star counts turn out to be compatible with the data from major photometric surveys such as SDSS and 2MASS \citep[see][for details]{Girardi2005, Girardi2012}. As compared to observed stellar catalogues, {\sc trilegal} provides about the same star counts as a function of coordinates, magnitudes and colours, but also additional information such as the evolutionary stage, mass, age, radius and distance. Stellar properties can be straightforwardly translated into  reasonable predictions of average or global asteroseismic parameters \citep[e.g. see][]{Chaplin2013}: the average large frequency separation (\deltanu) and the frequency of maximum oscillations power (\numax). The average separation scales to very good approximation as the
square root of the mean density of the star, i.e.  $\deltanu \propto \rho^{1/2}$; whilst
\numax\ has been found to scale with a combination of surface gravity
and effective temperature that also describes the dependence of the cut-off frequency for acoustic waves in an isothermal atmosphere,
i.e. $\numax \propto g T_{\rm eff}^{-1/2}$  \citep[e.g., see][for a recent review]{Belkacem2013}. 

The HR diagram of the synthetic population simulated with {\sc trilegal} in the PLATO field STEP08 is shown in Fig. \ref{fig:cmd}. 
\subsection{Predicting asteroseismic parameters and their detectability}
We follow the approach described in \citet{Mosser2017} based on the work by \citet{Mosser2009}, and explore the effect of varying the duration of the observations ($\tau$) and the apparent magnitude range  ($m_{\rm V}$) on the asteroseismic yields expected from the underlying stellar population (see also \citealt{Hekker2012}). Specifically, we quantify for each star in the synthetic population:
\begin{itemize}
\item whether solar-like oscillations are detectable,
\item the expected uncertainty on \numax\ and \deltanu,
\item our ability to measure gravity-mode period spacing (\deltaP), and hence to use it as a discriminant of evolutionary state \citep[e.g. see][]{Bedding2011},
\item whether rotationally split pulsation frequencies can be measured, and hence if information on the internal rotational profile can be inferred from the data.
\end{itemize}

Results of our simulations are presented in Figures \ref{fig:cmd}, \ref{fig:yields}, and \ref{fig:kde_duration}, and discussed in the following section.

\subsubsection{Detectability of the oscillations}
\label{sec:simulations}
By increasing the duration of the observational runs not only the overall number of stars for which oscillations are detected increases considerably (50k, 120k, 170k stars for a duration of \mbox{30 d}, \mbox{150 d}, and \mbox{2 yr}, respectively), but also larger areas of the HR diagram are covered by objects having seismic information (see Fig. \ref{fig:cmd}). 

Moreover, the duration of the observations sets an upper limit on the radius/luminosity of stars with measurable oscillation parameters (stars of larger radii have more closely spaced pulsation periods; for the largest stars, these periods become longer than the duration of the observations themselves). This also has implications on the distances that can be probed by such stars, for a given apparent magnitude. For instance, while the overall number of stars with detectable oscillations doubles when comparing yields from observations with durations of 150-d versus 30-d,, the number of stars at distances larger than 5 kpc becomes five times higher (here we are considering a lower brightness limit of $m_{\rm V}=15$). 

The lower limit on the intrinsic luminosity of stars with detectable oscillations becomes strongly dependent on the duration of the observations, especially for stars with apparent magnitudes $m_{\rm V} > 14$ (as illustrated by Figs \ref{fig:cmd} and \ref{fig:yields}), where the detectability is hampered by the increasing noise level (and by the intrinsically low pulsational amplitudes, which decrease with decreasing luminosity -- see e.g. \citealt{Kjeldsen1995, Samadi2012, Huber2010, Baudin2011}).

To account for the decreased detectability of solar-like oscillations in stars approaching the red edge of the classical pulsators instability strip we have followed the approach described in \citet{Chaplin2011}.  An in-depth study of the transition between solar-like and classical pulsations, also taking into account the effects of activity on the detectability of oscillation modes \citep[e.g., see][]{Garcia2010} is beyond the scope of this paper. 

Another fundamental detection limit is defined by the Nyquist frequency of the time series, which in this case, assuming a cadence of 600 s, is set to 833 $\mu$Hz\footnote{This limit does not apply to {low-mass} main-sequence and sub-giant stars which will be part of PLATO's core-target list, and which will be studied for asteroseismology using high-cadence data \citep[see][]{Rauer2014}. 
}, which is significantly higher than {\it Kepler}'s 278 $\mu$Hz. This opens the door to detecting oscillations in thousands of stars during their subgiant  phase ($\log{g} \simeq 3.5-4$, see Fig. \ref{fig:yields}). These objects are key to constraining transport processes of chemicals and the distribution (and evolution) of angular momentum inside stars \citep[e.g. see][]{Deheuvels2014}. 

As mentioned earlier, we have taken $m_{\rm V}=15$ to be the faint magnitude limit in the simulations. However, Fig. \ref{fig:yields} suggests that, provided  contamination from nearby sources is not severe, PLATO will be able to detect oscillations for fainter stars, at least if the duration of observations exceeds 30 days.

\subsubsection{Seismic parameters that can be measured from the spectra}
A more detailed description of what physical properties can be extracted from data of different durations can be inferred from Fig. \ref{fig:yields}. We notice that a measurement of gravity-mode period spacing is most useful, for population studies at least, in stars where a possible ambiguity in the evolutionary state is present ($\log{g} \sim 2.5$, see upper panel of Fig. \ref{fig:yields}). Our simulations show that, for such stars, a precise measurement of the period spacing is possible for observations of about 5 months or longer.

Even longer datasets are required if one aims at measuring rotationally-split frequencies in stars up to the core-He burning phase, which enables one to recover information about the internal rotational profile \citep[e.g. see][]{Beck2012, Deheuvels2012, Deheuvels2014, Mosser2012, Eggenberger2012}, or to infer the inclination of the star's rotational axis with respect to the line of sight \citep[e.g. see][]{Gizon2013,Chaplin2013b, Huber2013, Corsaro2017}.

The length of the observations strongly influences both the detection yields and the precision on the measurements of the average seismic parameters of solar-like oscillators, which affects the precision of the inferred stellar properties. In our simulations we have used data on \emph{Kepler} red-giant stars to quantify the uncertainties on \deltanu\ and \numax\  (see \citealt{Mosser2017}) for stars in the synthetic population. These uncertainties account for an irreducible limit in precision. It is about \deltanu/200 for \deltanu\ (dominated by the intrinsic variation in  \deltanuF\ as a function of mode frequency mainly due to acoustic glitches, e.g. see \citealt{Miglio2010, Mazumdar2012, Vrard2015}), and it is about \deltanu/5 for \numax\ (predominantly due to stochastic excitation and damping leading to intrinsic variability of the shape of the oscillation excess power). Estimating how the uncertainties on the measured seismic properties map onto the precision of the inferred stellar properties (primarily mass, hence age) is discussed in the next Section.  We notice that the uncertainties resulting from the simulations adopted here agree with the results from the approach presented in \citet{Davies2016}, where the seismic parameters determined  from varying the length of the time series representing different space missions have been compared in a case study based on a specific star.

\begin{figure}
\includegraphics[width=\linewidth]{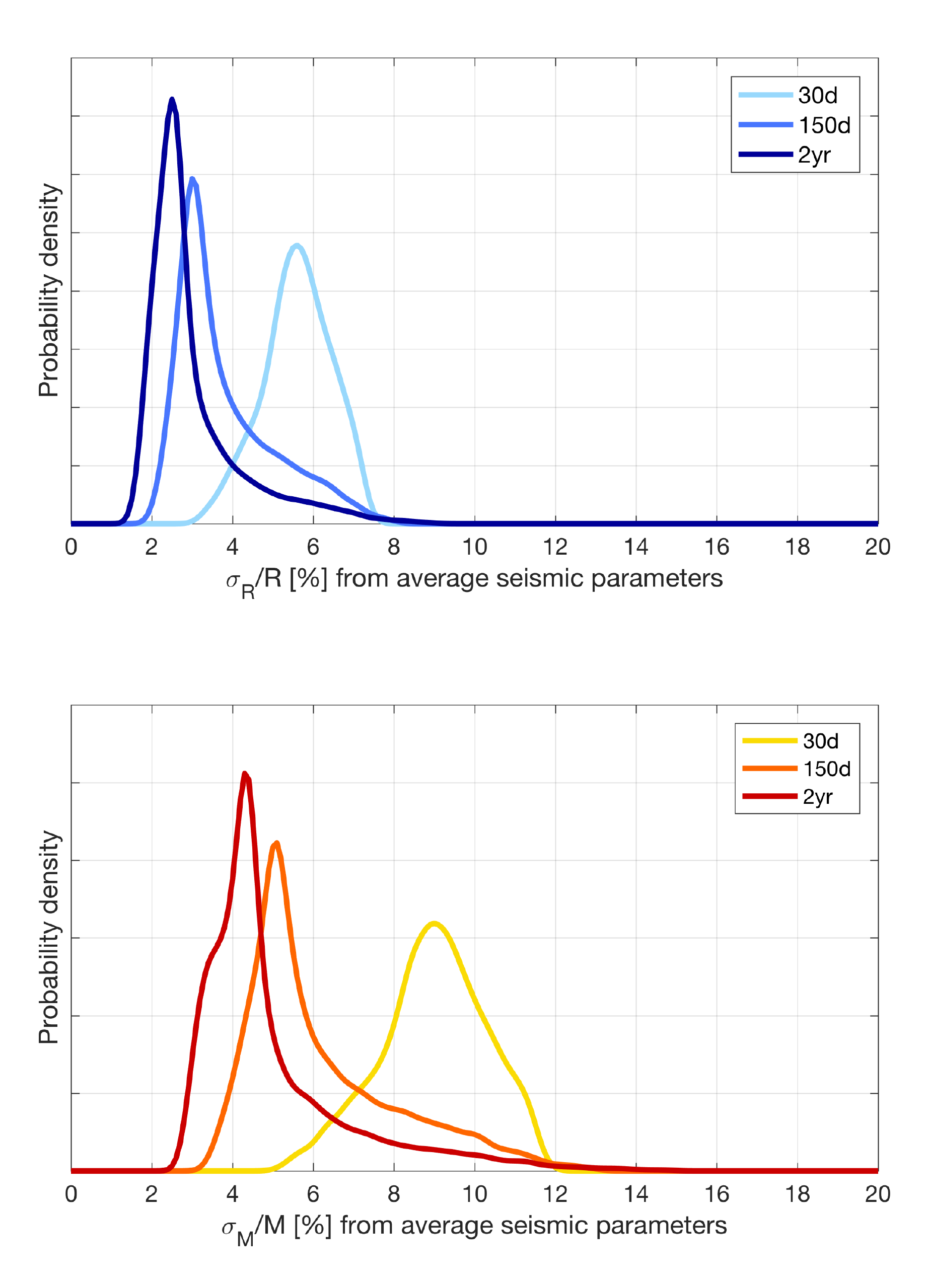}
\caption{Distribution of the expected precision on radius ({\it upper panel}) and mass ({\it lower panel}) for stars with detectable oscillations (see Fig. \ref{fig:yields}). The three lines in each panel show the effect of increasing the duration of the observations, from 30 d to 2 yr. Masses and radii are determined by combining \deltanu, \numax, and $T_{\rm eff}$ and their uncertainties.}
\label{fig:kde_duration}
\end{figure}

\begin{figure}
\includegraphics[width=\linewidth]{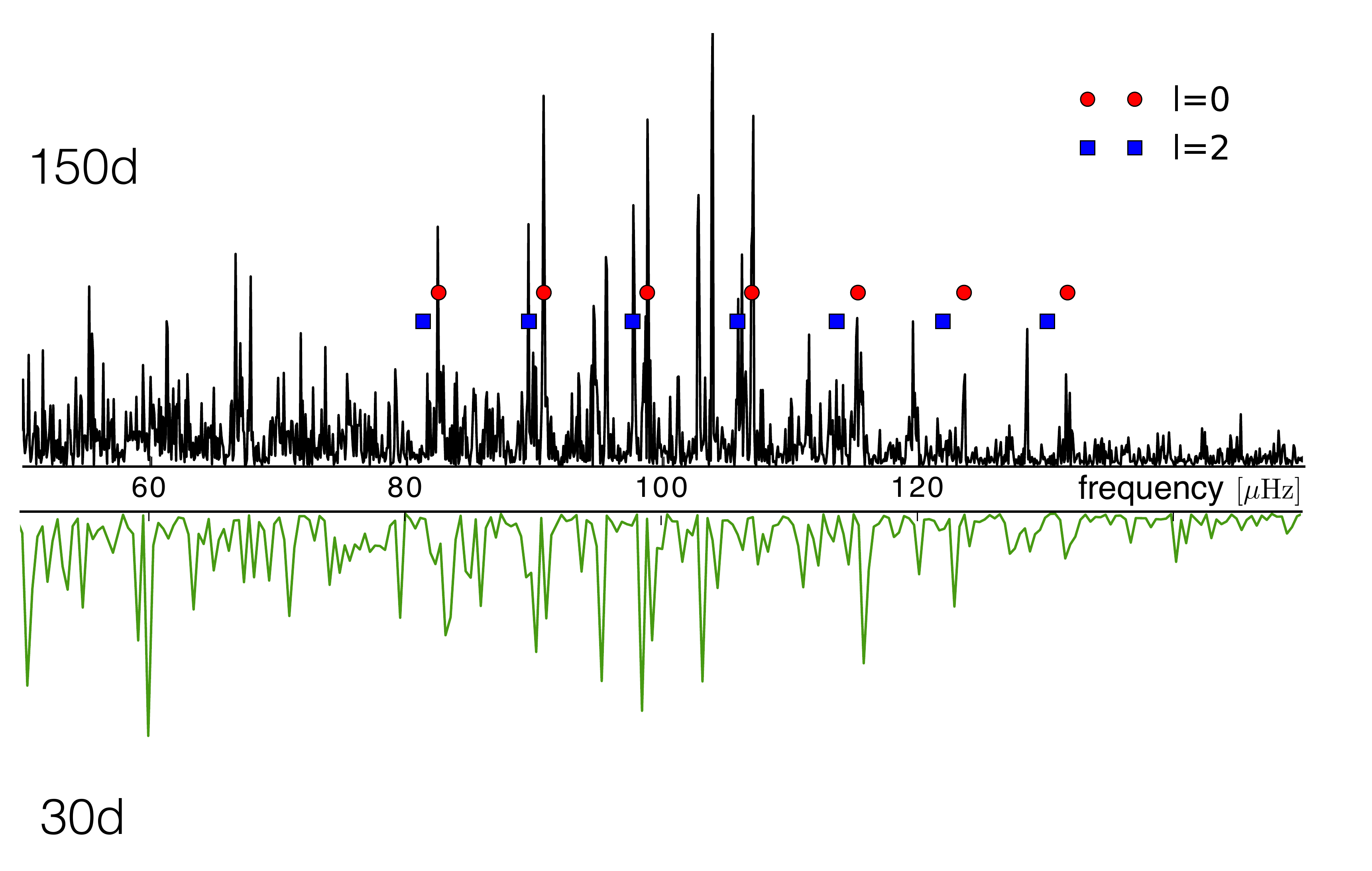}
\caption{Power spectral density as a function of frequency obtained by considering time series of duration 150 d and 30 d for a bright  ($m_{\rm V}=9$) giant observed by {\it Kepler}. The individual mode frequencies of radial (l=0) and quadrupolar modes (l=2) are indicated in the upper panel by red circles and blue squares, respectively. In the 30-d-long time series the robust identification of individual mode frequencies is hindered by the low frequency resolution, leading to a much reduced precision and accuracy on the inferred properties of the individual modes.}
\label{fig:spectra_duration}
\end{figure}

\begin{figure*}
\begin{center}
\includegraphics[width=.95\linewidth]{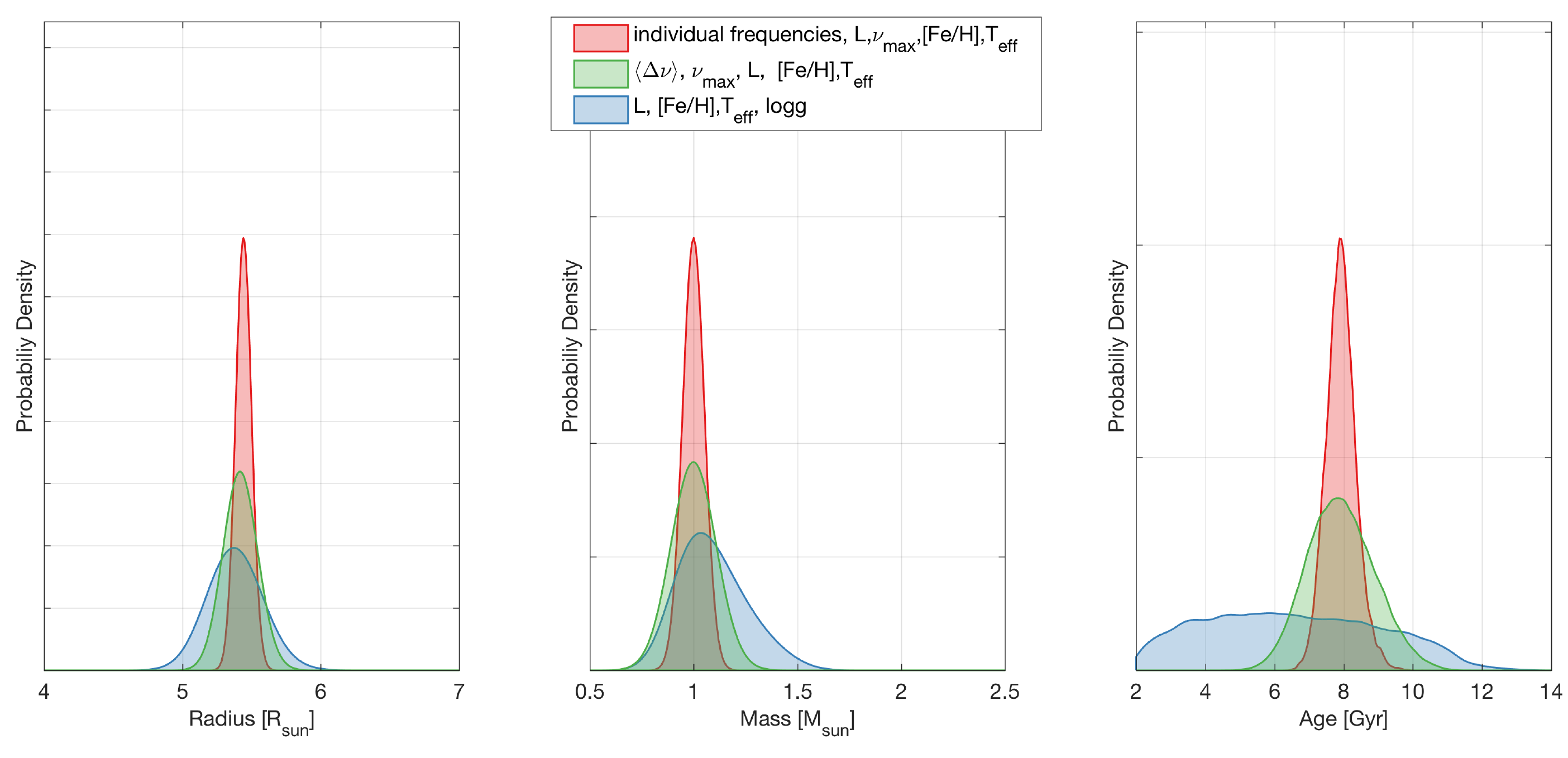}
\caption{Posterior probability density function of radius ({\it left panel}), mass ({\it middle panel}), and age ({\it right panel}) obtained by considering different combinations of seismic, spectroscopic, and astrometric constraints for a bright ($m_{\rm V}=9$) RGB star with \numax$\sim 110\, \mu$Hz. The assumed length of the observations is 150 d. The use of individual mode frequencies is expected to provide significantly improved precision  on the inferred stellar properties (see text for details).}
\label{fig:sigma_stellarprop}
\end{center}
\end{figure*}

\subsubsection{Frequencies of individual pulsation modes}
While average seismic parameters provide very useful estimates of global stellar properties, the highest levels of precision and accuracy are obtained when comparing observed individual frequencies to stellar models.

To assess the impact on the inferred stellar properties of the ability of measuring individual mode frequencies, we have considered a typical red-giant star observed by {\it Kepler}, with \numax $\sim 110$ $\mu$Hz,  and  divided up its time series into segments of different duration. Following the approach described in \citet{Davies2016b}, we then determined individual-mode frequencies  and their uncertainties. We have considered a star sufficiently bright so that the dominant source of background noise across the region occupied by the modes in the frequency-power spectrum is of stellar origin. As shown, e.g., by comparing a spectrum resulting from a 30-d to 150-d-long observations (see Fig. \ref{fig:spectra_duration}), a shorter length of the observations leads to a lower resolution of the power spectrum, making it harder to identify radial modes in the complex (and degraded) frequency spectrum. 

We limited the analysis to radial-mode frequencies, and find that, for all but the shortest time series ($\tau=30$ d) it was possible to determine individual-mode frequencies, albeit with complications related to disentangling radial modes from the more complex pattern of dipolar and quadrupolar modes  (typically for $\tau<150$ d). We have then focussed on the 150-d-long time series which led to uncertainties $\sigma_\nu$ with values in the range $0.04-0.09$ $\mu$Hz for the seven radial modes detected, which thus have a typical relative precision of the order of $10^{-4}$ -- $10^{-3}$.

\subsection{Mapping anticipated seismic constraints onto precision of the inferred stellar properties}
\label{sec:star_prop}
First, we assume that the available constraints are average seismic parameters (\deltanu\ and \numax) and $T_{\rm eff}$. Examples of how the expected precision of radius and mass depends on the duration of the observations are presented in Fig. \ref{fig:kde_duration}.  Although ages (and their uncertainties) cannot be inferred directly from seismic scaling relations,  the uncertainty on the age is expected to be indicatively a factor 3 larger than that on mass, based on the tight mass-age relation illustrated in Fig. \ref{fig:agescale}.

This means that 30-d-long observations would restrain our ability to infer ages to $\sim 40\%$, which is comparable to what one would expect for nearby stars without seismic constraints.  On the other hand, the 150-d-long time series would lead to a two-fold improvement in the precision. 
For a more in-depth description on how the expected uncertainties on stellar radius, mass, and age for red-giant stars depend on the assumed constraints and on (some of) the uncertainties in the models we refer to e.g. \citet{Noels2015, Noels2016, Casagrande2016, Rodrigues2017,Rendle2017}. 

In contrast to the case of {solar-like pulsating} main-sequence stars  (see e.g. \citealt{Lebreton2014,SilvaAguirre2015}), the effectiveness of individual mode frequencies in determining stellar properties for giant stars has yet to be fully explored. 
However, when individual mode frequencies are available as additional constraints, expectations are that the precision and accuracy on the inferred radii and masses (hence age) of red giants are significantly improved \citep{Huber2013, Perez2016, Lillo-Box2014}. To illustrate the expected gain in \emph{precision} when including radial-mode frequencies, in Fig. \ref{fig:sigma_stellarprop} we show the posterior probability distribution functions for mass and age which we obtain by including different sets of constraints in the inference procedure. The example is limited to the 150-d-long case and shows the expected precision on a typical low-luminosity RGB star. 
We have assumed a length of the observations of 150 d,  which would allow us to clearly identify modes, and determine the evolutionary state, and we have thus taken uncertainties on seismic parameters resulting from the simulations described above. 

We have then run the modelling pipeline {\sc aims} \citep{Reese2016, Rendle2017} that enables statistically robust inference on stellar properties, crucially including as constraints individual mode frequencies, and compared the posterior probability distribution functions of radius, mass, and age assuming astrometric and spectroscopic constraints only ($T_{\rm eff}$, [Fe/H], $\log{g}$, and luminosity as expected from Gaia for a nearby star), and then adding either average seismic constraints (\deltanu, \numax) or  individual radial mode frequencies.  {We assumed the following uncertainties on non-seismic constraints: 0.2 dex in \logg, 0.15 dex in [Fe/H] and  3\%  in luminosity (see \citealt{Rodrigues2017} for additional tests).}

\emph{When compared to the case of spectroscopic and astrometric constraints only, one can expect a 2.5-fold improvement in the precision on age when adding average seismic constraints, and  a dramatic 6-fold improvement when one is able to make use of the much more precise individual radial mode frequencies.} 
Data of such quality would thus make it possible to reach the desired precision in age ($\lesssim 10\,\%$). For the more ambitious goal to achieve similar level of accuracy, one would have to couple these data with stringent tests of models of stellar structure and evolution, which is one of the core science aims of the PLATO mission.

\subsubsection{Distances and interstellar reddening}
\label{sec:distance}
\begin{figure}
 \includegraphics[width=\linewidth]{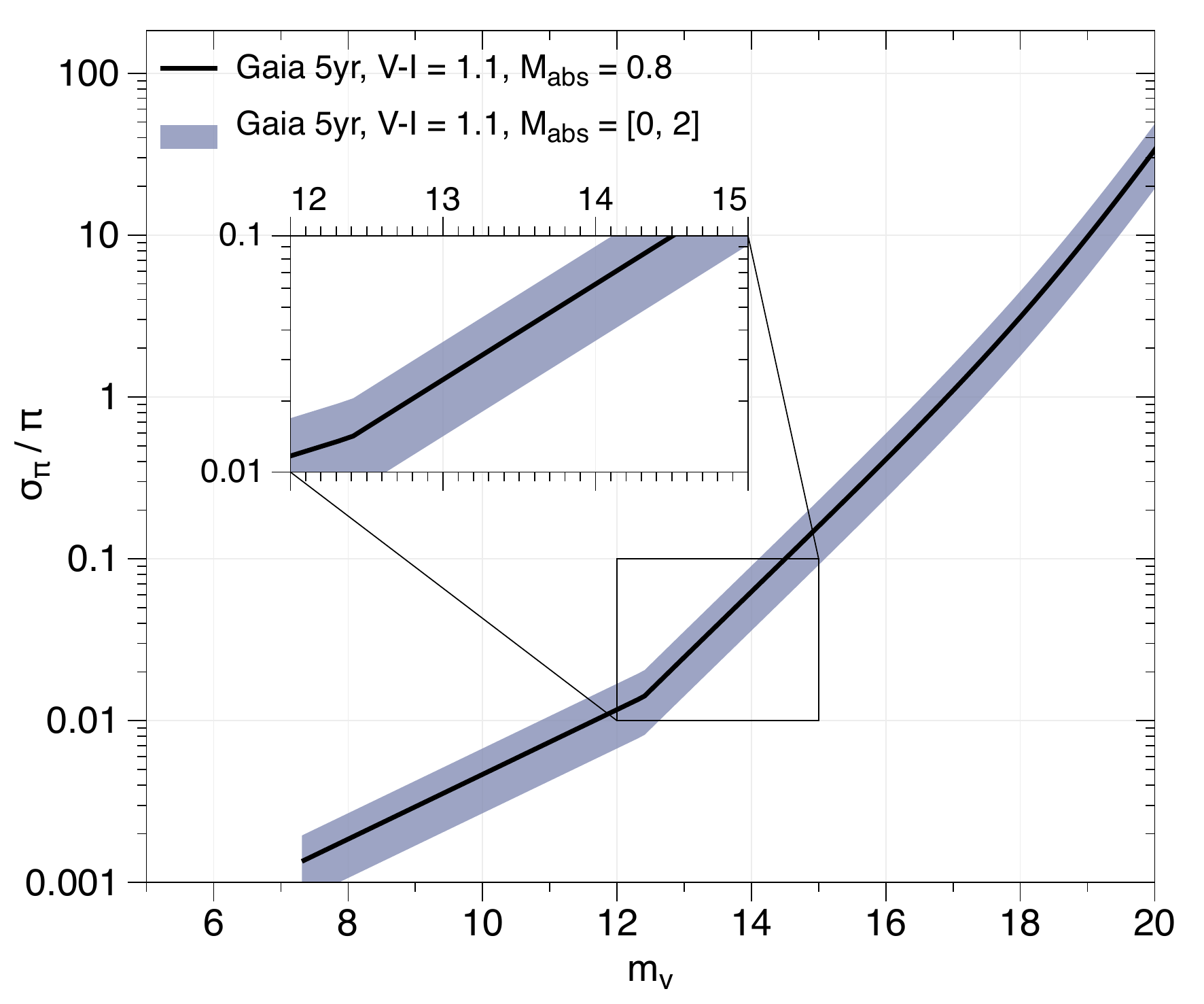}
 \caption{Gaia's  end-of-mission relative parallax error, evaluated for typical red-giant stars with detectable solar-like oscillations. The solid line illustrates the case of stars with an absolute $V$-band magnitude representative of red-clump stars. The precision achieved by seismology (few percent) is comparable or better than Gaia's for stars fainter than $m_{\rm V}=13-14$ (i.e. distances $\gtrsim 3$ kpc, see also \citealt{Huber2017}). Gaia parallax performance estimate adapted from: \texttt{\scriptsize https://www.cosmos.esa.int/web/gaia/science-performance} (using astrometric error model of \citealt{LL:GAIA-JDB-022} and colour transformation of \citealt{2010A&A...523A..48J}).}
 \label{fig:gaia}
 \end{figure}

\begin{figure}
\includegraphics[width=\linewidth]{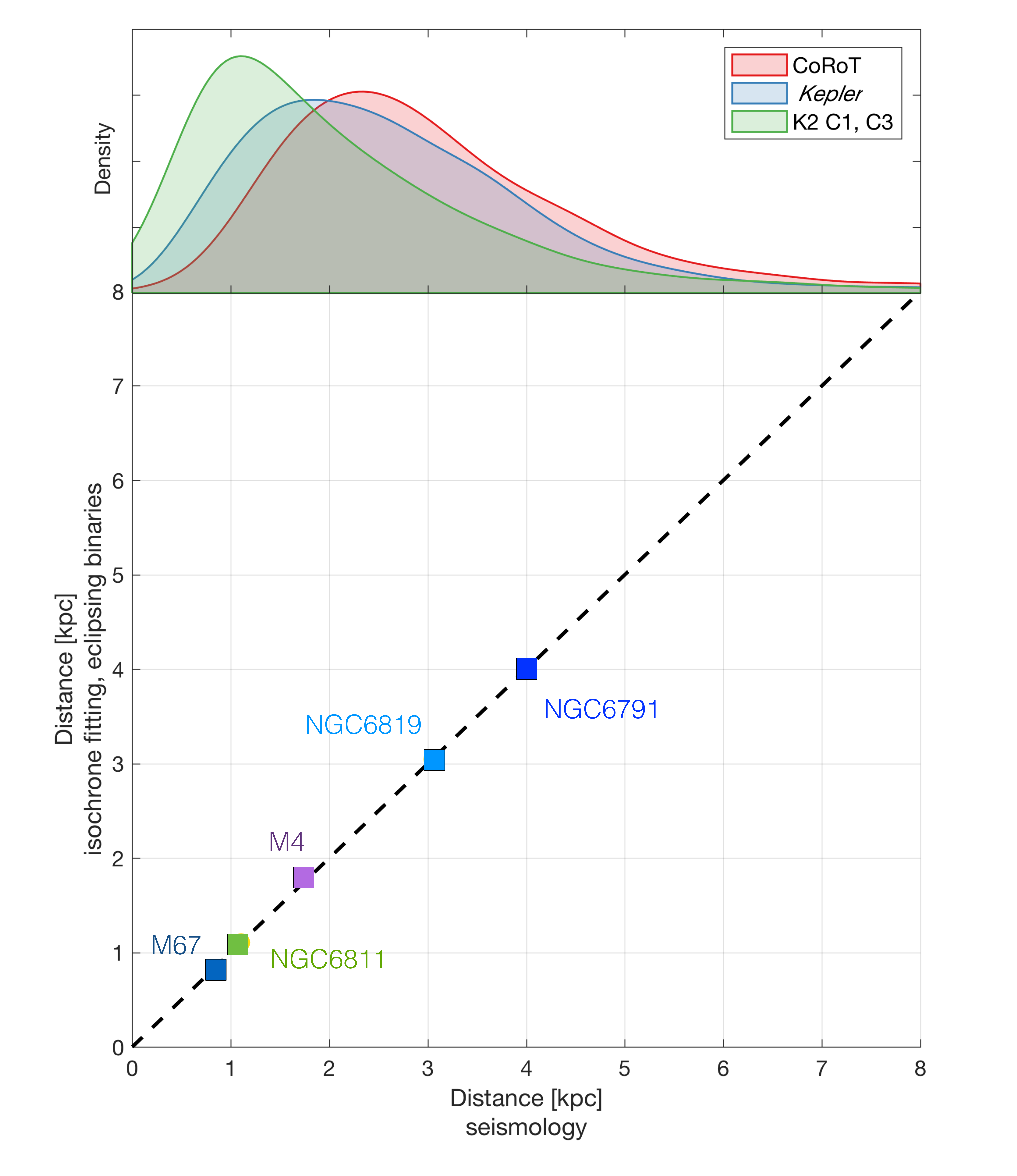}
\caption{{\it Lower panel:} asteroseismic distance scale for solar-like oscillating giants, presenting a comparison between seismic distances against benchmark distances of clusters, the latter obtained via isochrone fitting and/or based on eclipsing binaries. Distances are taken from \citet{Brogaard2016,Miglio2016,Stello2016,Handberg2017, Sandquist2016, Molenda2014}. {\it Upper panel:} Distribution of distances for targets in various asteroseismic missions. The different duration of the observations, coupled with the mission-specific target selection function explain the different distributions. Longer observations allow the measurement of oscillations in longer-period (hence, in general, intrinsically brighter and more distant) stars. CoRoT, in its so-called ``exo-field'', targeted stars fainter than {\it Kepler}. }
\label{fig:dist_scale}
\end{figure}

Seismic constraints can be combined with effective temperature and apparent photometric magnitudes to determine distances (see Fig. \ref{fig:dist_scale} and \citealt{Miglio2013,Rodrigues2014,Mathur2016,Anders2017a}). 
Such distances typically reach a level of precision of few percent ($2-5\,\%$, depending on the duration of the observations, see e.g. \citealt{Rodrigues2014}). Similarly to period-luminosity relations for classical pulsators, their precision depends little on the distance itself, as long as a robust detection of the oscillations is achievable.
Consequently, seismic distances will have comparable if not superior precision to Gaia for stars with $m_{\rm V} \gtrsim 13$, i.e. giant stars beyond $\sim 3$ kpc (see Fig. \ref{fig:gaia} and \citealt{Huber2017}). One could thus select targets to ensure that PLATO can also significantly improve the cartography of the Milky Way, given that oscillations are expected to be detectable for significantly fainter magnitudes (see Section \ref{sec:simulations} and Fig. \ref{fig:yields}). We note also that the prime targets for PLATO, i.e. bright stars, will play a fundamental role in testing the accuracy of the seismic distance scale, benefiting from negligible extinction, exquisite seismic, spectroscopic, photometric, astrometric data, and for some targets, interferometric constraints \citep[see e.g.][]{SilvaAguirre2012, Huber2012, Lagarde2015}.

Moreover, as a byproduct of the analysis,  3-D reddening maps can be determined by fitting the spectral-energy distributions in several photometric bands, and combining them with spectroscopic effective temperatures and precise bolometric luminosities from seismology \citep[see][for a detailed description of the method]{Rodrigues2014}. 

\subsubsection{Synergies with spectroscopic surveys}
\label{sec:spectroscopy}

An additional stellar property that asteroseismic constraints can deliver with high precision is surface gravity ($\sigma_{\log{g}} \lesssim 0.05$ dex, see e.g. \citealt{Morel2015}, and references therein).
Given the difficulties associated with measuring $\log{g}$ via spectroscopic analyses, large-scale spectroscopic surveys have now included solar-like oscillating stars among their targets, as key calibrators of surface gravity. For instance, CoRoT targets are now being observed by the Gaia-ESO Survey \citep{Valentini2016,Pancino2017}, APOGEE \citep{Anders2017a}, and GALAH \citep{Martell2017}. \Kepler\ targets have been used for calibrating stellar surface gravities in APOGEE \citep{Pinsonneault2014} and LAMOST \citep{Wang2016}.

\begin{figure}
\includegraphics[width=\linewidth]{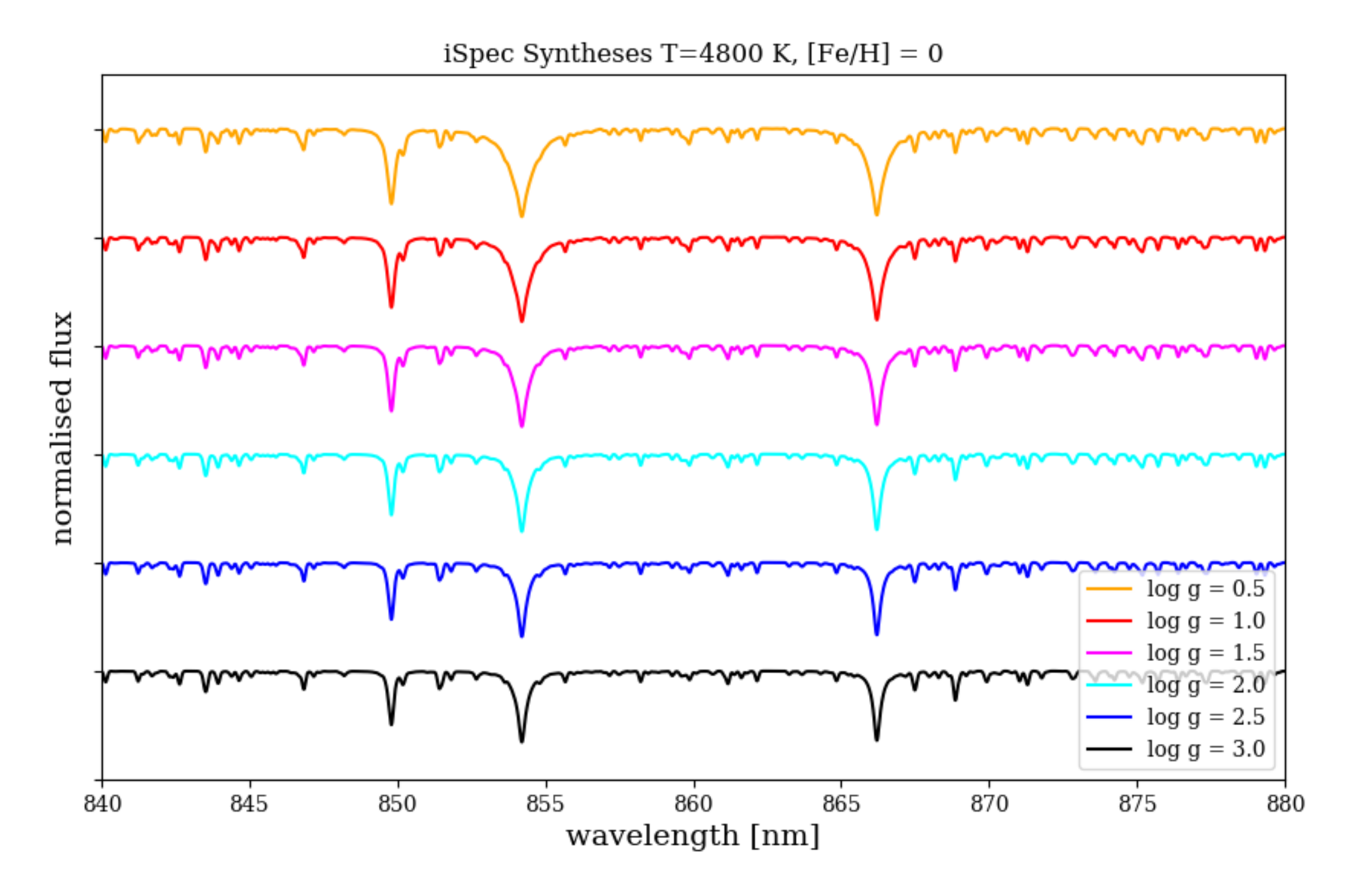}
\caption{Synthetic spectra of a solar-metallicity star with \mbox{$T_{\rm eff}=4800$ K} in the wavelength range and resolution of RAVE. Each line represents a spectrum with different  $\log{g}$, from 0.5 to 3.0 in steps of 0.5 dex, showing how little spectral features change with surface gravity. Spectra are synthesised using the iSpec package \citep{spec}, considering the Turbospectrum \citep{turbospectrum} radiative transfer code and MARCS \citep{MARCS} atmosphere models.}
\label{fig:spectra_logg}
\end{figure}

Recently, K2 targets at different locations \citep[e.g. see][]{Howell2014,Stello2015} have become the key stars for cross-calibrating several surveys.  
An  example of the impact of having seismic surface gravities for several stars included in spectroscopic surveys has been recently shown for RAVE.
The RAVE survey collected intermediate resolution spectra  
around the Ca triplet. 
This wavelength interval, despite being excellent for deriving radial velocities, contains few spectral lines 
resulting in degeneracies of stellar parameters: lines produced in stars with different surface gravities and at the same temperature are hardly discernible, as illustrated in Fig.~\ref{fig:spectra_logg}. 
K2 observed 87 RAVE red giants during Campaign 1, and the seismically inferred surface gravity provided a calibration for the \logg\ for giants \citep{Valentini2017}. Abundances have been recomputed then using these newly {\it calibrated} gravities, and presented in the RAVE-DR5-SC catalog \citep{Kunder2017}.

Additionally, beyond improving stellar parameters derived from spectra, seismic information has become critical in the era of high-precision chemical abundance determination analyses. 
In a new approach, where atmospheric parameters are computed by {\it fixing} \logg\ to the seismic value, and iteratively deriving the surface temperature and overall metallicity
[M/H] (thus ensuring high consistency among all stellar parameters), \citet{Morel2014} and \citet{Valentini2016,Valentini2017} have demonstrated that higher accuracy on chemical abundances can be achieved.

A further example was shown using high-resolution (R $\sim$ 22,500)
spectra from APOGEE, 
 in the $H$-band ($1.5-1.7 \mu$m). In this wavelength regime there is a lack of usable Fe II lines, which are widely used to constrain the surface gravity spectroscopically. Building on the work of \citet{Pinsonneault2014}, \citet{Hawkins2016} was able to show that using the seismic information (and adopting the APOGEE surface temperatures) one can {\it significantly} improve the precision and accuracy of stellar parameters and chemical abundances derived from APOGEE spectra. 
The seismic data for the APOGEE+\Kepler\ sample has also been used to identify the spectral regions that are most sensitive to \logg\ which can be used to find novel ways of constraining this difficult parameter beyond the standard Fe II ionization balance technique \citep[e.g.][]{Masseron2017}.
 
The spectroscopic follow-up of PLATO's targets by several planned large-scale surveys (e.g. 4MOST, WEAVE,  SDSS-V, see also Fig. \ref{fig:fields}) will not only be beneficial to the calibration of spectroscopic analysis procedures, but will allow for precise chemical abundance determinations which are key to inferring precise  stellar properties (in particular age),  to testing stellar models, and, notably, for informing models of Galactic chemical evolution and to help identify populations of stars with a common origin \citep[e.g. see][]{Freeman2002}. In particular, observing how individual star clusters have spread out is the most direct measure of radial migration with cosmic time  \citep{Bland-Hawthorn2010}. 

\section{Summary}
\label{sec:summary}
Deciphering the assembly history of the Milky Way is a formidable task, which becomes possible only if one can produce high resolution chrono-chemo-kinematical maps of the Galaxy. 

Currently a wealth of data is being gathered on ensembles of stars with the aim of improving our knowledge of the Milky Way structure and of its chemodynamical properties. The ESA Gaia satellite, with its second data release, will soon deliver an accurate 3-D map and proper motions of all detected stars, and radial velocities for bright stars throughout our Galaxy. Additional crucial information, both on velocities and chemical abundances, will come from several ongoing/planned spectroscopic surveys such as RAVE, SEGUE, APOGEE, Gaia-ESO, LAMOST, GALAH, WEAVE and 4MOST. These data will soon provide us with a well-defined view on the current chemo-kinematical structure of the Milky Way, but will only enable a blurred view on the temporal sequence that led to the present-day Galaxy. The framework for chemodynamical models tailored to the Milky Way now exists \citep[e.g.][]{Minchev2014}, as well as tools to best compare model predictions to the data  \cite[e.g.][]{Sharma2011, Anders2016}.

Astrometric and spectroscopic constraints alone will not enable precise and accurate estimates of stellar age. This is particularly true for red giant stars, which are the primary tracers of the Milky Way's structure. Asteroseismology clearly provides the way forward: solar-like oscillating giants are excellent clocks thanks to the availability of seismic constraints on their mass and to the tight age-initial-mass relation they adhere to. The potential of asteroseismology for constraining evolutionary models of the Milky Way has now been demonstrated thanks to the ongoing exploitation of data from the pioneering photometric missions CoRoT, {\it Kepler}, and K2.  

These missions, however, are limited in either Galactic volume coverage or duration of the observations, which limits the precision one can achieve on the inferred stellar properties, chiefly age.  In this paper we have identified five key  questions (see Section \ref{sec:openquestions}) that we believe will need precise and accurate ages for large samples of stars to be addressed, and we identified the requirements in terms of number of targets and  the precision on the stellar properties that are needed to tackle such questions (Section \ref{sec:desiderata}).

By quantifying the seismic yields expected from PLATO, we have shown in Section \ref{sec:seismo} that the requirements outlined in Section \ref{sec:desiderata} are  within the capabilities of the current PLATO design, provided that observations are sufficiently long to identify the evolutionary state and allow robust and precise determination of acoustic-mode frequencies.  This will allow us to harvest data of sufficient quality to reach a 10\% precision in age. This is a fundamental prerequisite to then reach the more ambitious goal of a similar level of accuracy, which will only be possible if coupled with a careful appraisal of systematic uncertainties on age deriving from our limited understanding of stellar physics; a goal which conveniently falls within the main aims of PLATO's core science.
\emph{We therefore strongly endorse PLATO's current design and proposed observational strategy, and conclude that PLATO, {\it as it is}, will be a legacy mission for Galactic archaeology.}

\newpage
\acknowledgements
We are extremely grateful to the International Space Science Institute (ISSI) for support provided to the \mbox{asterosSTEP} ISSI International Team\footnote{\texttt{http://www.issibern.ch/teams/asterostep/}}$^,$\footnote{\texttt{http://www.asterostep.eu/}}.
This article made use of AIMS, a software for fitting stellar pulsation data, developed in the context of the SPACEINN network, funded by the European Commission's Seventh Framework Programme. This research has made use of ``Aladin sky atlas'' developed at CDS, Strasbourg Observatory, France. 
CC acknowledges support from DFG Grant CH1188/2-1 and from the ChETEC COST Action (CA16117), supported by COST (European Cooperation in Science and Technology). Funding for the Stellar Astrophysics Centre is provided by The Danish National Research Foundation (Grant agreement no.: DNRF106). AM, GRD, WJC and IWR acknowledge the support of the UK Science and Technology Facilities Council (STFC). LG and TS acknowledge support from PRIN INAF 2014 -- CRA 1.05.01.94.05. PJ acknowledges European Union FP7 programme through ERC grant number 320360. LC gratefully acknowledges support from the Australian Research Council (grants DP150100250, FT160100402).
SB acknowledges NASA grant NNX16AI09G.
MC and PPA acknowledge support from FCT through the grant UID/FIS/04434/2013 and the contract IF/00894/2012/ and POPH/FSE (EC) by FEDER funding through the programme COMPETE. RAG acknowledges the support from CNES. L.Gizon acknowledges research funding from DLR under the PLATO Data Center and from the NYUAD Institute under grant no. G1502. NL thanks financial support from ``Programme National de Physique Stellaire'' (PNPS) and from ``Programme National Cosmologie et Galaxies'' of CNRS/INSU, France. NL acknowledges financial support from the CNES fellowship.  UH acknowledges support from the Swedish National Space Board (SNSB/Rymdstyrelsen). SM acknowledges support from the NASA grant NNX16AJ17G. PM and JM acknowledge support from the ERC Consolidator Grant funding scheme ({\em project STARKEY}, G.A. n. 615604). TM acknowledges financial support from Belspo for contract PRODEX PLATO. DS is the recipient of an Australian Research Council Future Fellowship (project number FT1400147).   AS acknowledges funding from ESP2015-66134-R (MINECO). VSA acknowledges support from VILLUM FONDEN (research grant 10118).

\bibliographystyle{an}
\bibliography{plato}

\begin{thebibliography}{177}
\expandafter\ifx\csname natexlab\endcsname\relax\def\natexlab#1{#1}\fi

\bibitem[{{Abadi} {et~al.}(2003){Abadi}, {Navarro}, {Steinmetz}, \&
  {Eke}}]{Abadi2003}
{Abadi}, M.~G., {Navarro}, J.~F., {Steinmetz}, M., \& {Eke}, V.~R. 2003, \apj,
  591, 499

\bibitem[{{Amores} {et~al.}(2017){Amores}, {Robin}, \& {Reyle}}]{Amores2017}
{Amores}, E.~B., {Robin}, A.~C., \& {Reyle}, C. 2017, ArXiv e-prints

\bibitem[{{Anders} {et~al.}(2017{\natexlab{a}}){Anders}, {Chiappini},
  {Minchev}, {Miglio}, {Montalb{\'a}n}, {Mosser}, {Rodrigues}, {Santiago},
  {Baudin}, {Beers}, {da Costa}, {Garc{\'{\i}}a},
  {Garc{\'{\i}}a-Hern{\'a}ndez}, {Holtzman}, {Maia}, {Majewski}, {Mathur},
  {Noels-Grotsch}, {Pan}, {Schneider}, {Schultheis}, {Steinmetz}, {Valentini},
  \& {Zamora}}]{Anders2017b}
{Anders}, F., {Chiappini}, C., {Minchev}, I., {et~al.} 2017{\natexlab{a}},
  \aap, 600, A70

\bibitem[{{Anders} {et~al.}(2017{\natexlab{b}}){Anders}, {Chiappini},
  {Rodrigues}, {Miglio}, {Montalb{\'a}n}, {Mosser}, {Girardi}, {Valentini},
  {Noels}, {Morel}, {Johnson}, {Schultheis}, {Baudin}, {de Assis Peralta},
  {Hekker}, {Theme{\ss}l}, {Kallinger}, {Garc{\'{\i}}a}, {Mathur}, {Baglin},
  {Santiago}, {Martig}, {Minchev}, {Steinmetz}, {da Costa}, {Maia}, {Allende
  Prieto}, {Cunha}, {Beers}, {Epstein}, {Garc{\'{\i}}a P{\'e}rez},
  {Garc{\'{\i}}a-Hern{\'a}ndez}, {Harding}, {Holtzman}, {Majewski},
  {M{\'e}sz{\'a}ros}, {Nidever}, {Pan}, {Pinsonneault}, {Schiavon},
  {Schneider}, {Shetrone}, {Stassun}, {Zamora}, \& {Zasowski}}]{Anders2017a}
{Anders}, F., {Chiappini}, C., {Rodrigues}, T.~S., {et~al.} 2017{\natexlab{b}},
  \aap, 597, A30

\bibitem[{{Anders} {et~al.}(2016){Anders}, {Chiappini}, {Rodrigues}, {Piffl},
  {Mosser}, {Miglio}, {Montalb{\'a}n}, {Girardi}, {Minchev}, {Valentini}, \&
  {Steinmetz}}]{Anders2016}
{Anders}, F., {Chiappini}, C., {Rodrigues}, T.~S., {et~al.} 2016, Astronomische
  Nachrichten, 337, 926

\bibitem[{{Anders} {et~al.}(2014){Anders}, {Chiappini}, {Santiago},
  {Rocha-Pinto}, {Girardi}, {da Costa}, {Maia}, {Steinmetz}, {Minchev},
  {Schultheis}, {Boeche}, {Miglio}, {Montalb{\'a}n}, {Schneider}, {Beers},
  {Cunha}, {Allende Prieto}, {Balbinot}, {Bizyaev}, {Brauer}, {Brinkmann},
  {Frinchaboy}, {Garc{\'{\i}}a P{\'e}rez}, {Hayden}, {Hearty}, {Holtzman},
  {Johnson}, {Kinemuchi}, {Majewski}, {Malanushenko}, {Malanushenko},
  {Nidever}, {O'Connell}, {Pan}, {Robin}, {Schiavon}, {Shetrone}, {Skrutskie},
  {Smith}, {Stassun}, \& {Zasowski}}]{Anders2014}
{Anders}, F., {Chiappini}, C., {Santiago}, B.~X., {et~al.} 2014, \aap, 564,
  A115

\bibitem[{{Arentoft} {et~al.}(2017){Arentoft}, {Brogaard}, {Jessen-Hansen},
  {Silva Aguirre}, {Kjeldsen}, {Mosumgaard}, \& {Sandquist}}]{Arentoft2017}
{Arentoft}, T., {Brogaard}, K., {Jessen-Hansen}, J., {et~al.} 2017, ArXiv
  e-prints

\bibitem[{{Athanassoula} {et~al.}(2017){Athanassoula}, {Rodionov}, \&
  {Prantzos}}]{Athanassoula2017}
{Athanassoula}, E., {Rodionov}, S.~A., \& {Prantzos}, N. 2017, \mnras, 467, L46

\bibitem[{{Babusiaux}(2016)}]{Babusiaux2016}
{Babusiaux}, C. 2016, \pasa, 33, e026

\bibitem[{{Baglin} {et~al.}(2006){Baglin}, {Auvergne}, {Barge}, {Deleuil},
  {Catala}, {Michel}, {Weiss}, \& {COROT Team}}]{Baglin2006}
{Baglin}, A., {Auvergne}, M., {Barge}, P., {et~al.} 2006, in ESA Special
  Publication, Vol. 1306, The CoRoT Mission Pre-Launch Status - Stellar
  Seismology and Planet Finding, ed. M.~{Fridlund}, A.~{Baglin}, J.~{Lochard},
  \& L.~{Conroy}, 33

\bibitem[{{Baudin} {et~al.}(2011){Baudin}, {Barban}, {Belkacem}, {Hekker},
  {Morel}, {Samadi}, {Benomar}, {Goupil}, {Carrier}, {Ballot}, {Deheuvels}, {De
  Ridder}, {Hatzes}, {Kallinger}, \& {Weiss}}]{Baudin2011}
{Baudin}, F., {Barban}, C., {Belkacem}, K., {et~al.} 2011, \aap, 529, A84

\bibitem[{{Beck} {et~al.}(2012){Beck}, {Montalban}, {Kallinger}, {De Ridder},
  {Aerts}, {Garc{\'{\i}}a}, {Hekker}, {Dupret}, {Mosser}, {Eggenberger},
  {Stello}, {Elsworth}, {Frandsen}, {Carrier}, {Hillen}, {Gruberbauer},
  {Christensen-Dalsgaard}, {Miglio}, {Valentini}, {Bedding}, {Kjeldsen},
  {Girouard}, {Hall}, \& {Ibrahim}}]{Beck2012}
{Beck}, P.~G., {Montalban}, J., {Kallinger}, T., {et~al.} 2012, \nat, 481, 55

\bibitem[{{Bedding} {et~al.}(2011){Bedding}, {Mosser}, {Huber},
  {Montalb{\'a}n}, {Beck}, {Christensen-Dalsgaard}, {Elsworth},
  {Garc{\'{\i}}a}, {Miglio}, {Stello}, {White}, {De Ridder}, {Hekker}, {Aerts},
  {Barban}, {Belkacem}, {Broomhall}, {Brown}, {Buzasi}, {Carrier}, {Chaplin},
  {di Mauro}, {Dupret}, {Frandsen}, {Gilliland}, {Goupil}, {Jenkins},
  {Kallinger}, {Kawaler}, {Kjeldsen}, {Mathur}, {Noels}, {Silva Aguirre}, \&
  {Ventura}}]{Bedding2011}
{Bedding}, T.~R., {Mosser}, B., {Huber}, D., {et~al.} 2011, \nat, 471, 608

\bibitem[{{Belkacem} {et~al.}(2013){Belkacem}, {Samadi}, {Mosser}, {Goupil}, \&
  {Ludwig}}]{Belkacem2013}
{Belkacem}, K., {Samadi}, R., {Mosser}, B., {Goupil}, M.-J., \& {Ludwig}, H.-G.
  2013, in Astronomical Society of the Pacific Conference Series, Vol. 479,
  Progress in Physics of the Sun and Stars: A New Era in Helio- and
  Asteroseismology, ed. H.~{Shibahashi} \& A.~E. {Lynas-Gray}, 61

\bibitem[{{Bensby} {et~al.}(2017){Bensby}, {Feltzing}, {Gould}, {Yee},
  {Johnson}, {Asplund}, {Mel{\'e}ndez}, {Lucatello}, {Howes}, {McWilliam},
  {Udalski}, {Szyma{\'n}ski}, {Soszy{\'n}ski}, {Poleski}, {Wyrzykowski},
  {Ulaczyk}, {Koz{\l}owski}, {Pietrukowicz}, {Skowron}, {Mr{\'o}z}, {Pawlak},
  {Abe}, {Asakura}, {Bhattacharya}, {Bond}, {Bennett}, {Hirao}, {Nagakane},
  {Koshimoto}, {Sumi}, {Suzuki}, \& {Tristram}}]{Bensby2017}
{Bensby}, T., {Feltzing}, S., {Gould}, A., {et~al.} 2017, ArXiv e-prints

\bibitem[{{Bensby} {et~al.}(2014){Bensby}, {Feltzing}, \& {Oey}}]{Bensby2014}
{Bensby}, T., {Feltzing}, S., \& {Oey}, M.~S. 2014, \aap, 562, A71

\bibitem[{{Bergemann} {et~al.}(2014){Bergemann}, {Ruchti}, {Serenelli},
  {Feltzing}, {Alves-Brito}, {Asplund}, {Bensby}, {Gruyters}, {Heiter},
  {Hourihane}, {Korn}, {Lind}, {Marino}, {Jofre}, {Nordlander}, {Ryde},
  {Worley}, {Gilmore}, {Randich}, {Ferguson}, {Jeffries}, {Micela},
  {Negueruela}, {Prusti}, {Rix}, {Vallenari}, {Alfaro}, {Allende Prieto},
  {Bragaglia}, {Koposov}, {Lanzafame}, {Pancino}, {Recio-Blanco}, {Smiljanic},
  {Walton}, {Costado}, {Franciosini}, {Hill}, {Lardo}, {de Laverny}, {Magrini},
  {Maiorca}, {Masseron}, {Morbidelli}, {Sacco}, {Kordopatis}, \& {Tautvai{\v
  s}ien{\.e}}}]{Bergemann2014}
{Bergemann}, M., {Ruchti}, G.~R., {Serenelli}, A., {et~al.} 2014, \aap, 565,
  A89

\bibitem[{{Bird} {et~al.}(2013){Bird}, {Kazantzidis}, {Weinberg}, {Guedes},
  {Callegari}, {Mayer}, \& {Madau}}]{Bird2013}
{Bird}, J.~C., {Kazantzidis}, S., {Weinberg}, D.~H., {et~al.} 2013, \apj, 773,
  43

\bibitem[{{Blanco-Cuaresma} {et~al.}(2014){Blanco-Cuaresma}, {Soubiran},
  {Heiter}, \& {Jofr{\'e}}}]{spec}
{Blanco-Cuaresma}, S., {Soubiran}, C., {Heiter}, U., \& {Jofr{\'e}}, P. 2014,
  \aap, 569, A111

\bibitem[{{Bland-Hawthorn} {et~al.}(2010){Bland-Hawthorn}, {Krumholz}, \&
  {Freeman}}]{Bland-Hawthorn2010}
{Bland-Hawthorn}, J., {Krumholz}, M.~R., \& {Freeman}, K. 2010, \apj, 713, 166

\bibitem[{{Boeche} {et~al.}(2014){Boeche}, {Siebert}, {Piffl}, {Just},
  {Steinmetz}, {Grebel}, {Sharma}, {Kordopatis}, {Gilmore}, {Chiappini},
  {Freeman}, {Gibson}, {Munari}, {Siviero}, {Bienaym{\'e}}, {Navarro},
  {Parker}, {Reid}, {Seabroke}, {Watson}, {Wyse}, \& {Zwitter}}]{Boeche2014}
{Boeche}, C., {Siebert}, A., {Piffl}, T., {et~al.} 2014, \aap, 568, A71

\bibitem[{{Boeche} {et~al.}(2013){Boeche}, {Siebert}, {Piffl}, {Just},
  {Steinmetz}, {Sharma}, {Kordopatis}, {Gilmore}, {Chiappini}, {Williams},
  {Grebel}, {Bland-Hawthorn}, {Gibson}, {Munari}, {Siviero}, {Bienaym{\'e}},
  {Navarro}, {Parker}, {Reid}, {Seabroke}, {Watson}, {Wyse}, \&
  {Zwitter}}]{Boeche2013}
{Boeche}, C., {Siebert}, A., {Piffl}, T., {et~al.} 2013, \aap, 559, A59

\bibitem[{{Borucki} {et~al.}(2010){Borucki}, {Koch}, {Basri}, {Batalha},
  {Brown}, {Caldwell}, {Caldwell}, {Christensen-Dalsgaard}, {Cochran},
  {DeVore}, {Dunham}, {Dupree}, {Gautier}, {Geary}, {Gilliland}, {Gould},
  {Howell}, {Jenkins}, {Kondo}, {Latham}, {Marcy}, {Meibom}, {Kjeldsen},
  {Lissauer}, {Monet}, {Morrison}, {Sasselov}, {Tarter}, {Boss}, {Brownlee},
  {Owen}, {Buzasi}, {Charbonneau}, {Doyle}, {Fortney}, {Ford}, {Holman},
  {Seager}, {Steffen}, {Welsh}, {Rowe}, {Anderson}, {Buchhave}, {Ciardi},
  {Walkowicz}, {Sherry}, {Horch}, {Isaacson}, {Everett}, {Fischer}, {Torres},
  {Johnson}, {Endl}, {MacQueen}, {Bryson}, {Dotson}, {Haas}, {Kolodziejczak},
  {Van Cleve}, {Chandrasekaran}, {Twicken}, {Quintana}, {Clarke}, {Allen},
  {Li}, {Wu}, {Tenenbaum}, {Verner}, {Bruhweiler}, {Barnes}, \&
  {Prsa}}]{Borucki2010}
{Borucki}, W.~J., {Koch}, D., {Basri}, G., {et~al.} 2010, Science, 327, 977

\bibitem[{{Bournaud}(2016)}]{Bournaud2016}
{Bournaud}, F. 2016, Galactic Bulges, 418, 355

\bibitem[{{Bournaud} {et~al.}(2009){Bournaud}, {Elmegreen}, \&
  {Martig}}]{Bournaud2009}
{Bournaud}, F., {Elmegreen}, B.~G., \& {Martig}, M. 2009, \apjl, 707, L1

\bibitem[{{Bovy} {et~al.}(2012){Bovy}, {Rix}, \& {Hogg}}]{Bovy2012}
{Bovy}, J., {Rix}, H.-W., \& {Hogg}, D.~W. 2012, \apj, 751, 131

\bibitem[{{Bressan} {et~al.}(2012){Bressan}, {Marigo}, {Girardi}, {Salasnich},
  {Dal Cero}, {Rubele}, \& {Nanni}}]{Bressan2012}
{Bressan}, A., {Marigo}, P., {Girardi}, L., {et~al.} 2012, \mnras, 427, 127

\bibitem[{{Brogaard} {et~al.}(2016){Brogaard}, {Jessen-Hansen}, {Handberg},
  {Arentoft}, {Frandsen}, {Grundahl}, {Bruntt}, {Sandquist}, {Miglio}, {Beck},
  {Thygesen}, {Kj{\ae}rgaard}, \& {Haugaard}}]{Brogaard2016}
{Brogaard}, K., {Jessen-Hansen}, J., {Handberg}, R., {et~al.} 2016,
  Astronomische Nachrichten, 337, 793

\bibitem[{{Brogaard} {et~al.}(2012){Brogaard}, {VandenBerg}, {Bruntt},
  {Grundahl}, {Frandsen}, {Bedin}, {Milone}, {Dotter}, {Feiden}, {Stetson},
  {Sandquist}, {Miglio}, {Stello}, \& {Jessen-Hansen}}]{Brogaard2012}
{Brogaard}, K., {VandenBerg}, D.~A., {Bruntt}, H., {et~al.} 2012, \aap, 543,
  A106

\bibitem[{{Brook} {et~al.}(2004){Brook}, {Kawata}, {Gibson}, \&
  {Freeman}}]{Brook2004}
{Brook}, C.~B., {Kawata}, D., {Gibson}, B.~K., \& {Freeman}, K.~C. 2004, \apj,
  612, 894

\bibitem[{{Cacciari} {et~al.}(2016){Cacciari}, {Pancino}, \&
  {Bellazzini}}]{Cacciari2016}
{Cacciari}, C., {Pancino}, E., \& {Bellazzini}, M. 2016, Astronomische
  Nachrichten, 337, 899

\bibitem[{{Campante} {et~al.}(2016){Campante}, {Lund}, {Kuszlewicz}, {Davies},
  {Chaplin}, {Albrecht}, {Winn}, {Bedding}, {Benomar}, {Bossini}, {Handberg},
  {Santos}, {Van Eylen}, {Basu}, {Christensen-Dalsgaard}, {Elsworth}, {Hekker},
  {Hirano}, {Huber}, {Karoff}, {Kjeldsen}, {Lundkvist}, {North}, {Silva
  Aguirre}, {Stello}, \& {White}}]{Campante2016}
{Campante}, T.~L., {Lund}, M.~N., {Kuszlewicz}, J.~S., {et~al.} 2016, \apj,
  819, 85

\bibitem[{{Casagrande} {et~al.}(2016){Casagrande}, {Silva Aguirre},
  {Schlesinger}, {Stello}, {Huber}, {Serenelli}, {Sch{\"o}nrich}, {Cassisi},
  {Pietrinferni}, {Hodgkin}, {Milone}, {Feltzing}, \&
  {Asplund}}]{Casagrande2016}
{Casagrande}, L., {Silva Aguirre}, V., {Schlesinger}, K.~J., {et~al.} 2016,
  \mnras, 455, 987

\bibitem[{{Casey} {et~al.}(2017){Casey}, {Hawkins}, {Hogg}, {Ness}, {Rix},
  {Kordopatis}, {Kunder}, {Steinmetz}, {Koposov}, {Enke}, {Sanders}, {Gilmore},
  {Zwitter}, {Freeman}, {Casagrande}, {Matijevi{\v c}}, {Seabroke},
  {Bienaym{\'e}}, {Bland-Hawthorn}, {Gibson}, {Grebel}, {Helmi}, {Munari},
  {Navarro}, {Reid}, {Siebert}, \& {Wyse}}]{Casey2017}
{Casey}, A.~R., {Hawkins}, K., {Hogg}, D.~W., {et~al.} 2017, \apj, 840, 59

\bibitem[{{Chaplin} {et~al.}(2011){Chaplin}, {Kjeldsen}, {Bedding},
  {Christensen-Dalsgaard}, {Gilliland}, {Kawaler}, {Appourchaux}, {Elsworth},
  {Garc{\'{\i}}a}, {Houdek}, {Karoff}, {Metcalfe}, {Molenda-{\.Z}akowicz},
  {Monteiro}, {Thompson}, {Verner}, {Batalha}, {Borucki}, {Brown}, {Bryson},
  {Christiansen}, {Clarke}, {Jenkins}, {Klaus}, {Koch}, {An}, {Ballot}, {Basu},
  {Benomar}, {Bonanno}, {Broomhall}, {Campante}, {Corsaro}, {Creevey}, {Esch},
  {Gai}, {Gaulme}, {Hale}, {Handberg}, {Hekker}, {Huber}, {Mathur}, {Mosser},
  {New}, {Pinsonneault}, {Pricopi}, {Quirion}, {R{\'e}gulo}, {Roxburgh},
  {Salabert}, {Stello}, \& {Suran}}]{Chaplin2011}
{Chaplin}, W.~J., {Kjeldsen}, H., {Bedding}, T.~R., {et~al.} 2011, \apj, 732,
  54

\bibitem[{{Chaplin} \& {Miglio}(2013)}]{Chaplin2013}
{Chaplin}, W.~J. \& {Miglio}, A. 2013, \araa, 51, 353

\bibitem[{{Chaplin} {et~al.}(2013){Chaplin}, {Sanchis-Ojeda}, {Campante},
  {Handberg}, {Stello}, {Winn}, {Basu}, {Christensen-Dalsgaard}, {Davies},
  {Metcalfe}, {Buchhave}, {Fischer}, {Bedding}, {Cochran}, {Elsworth},
  {Gilliland}, {Hekker}, {Huber}, {Isaacson}, {Karoff}, {Kawaler}, {Kjeldsen},
  {Latham}, {Lund}, {Lundkvist}, {Marcy}, {Miglio}, {Barclay}, \&
  {Lissauer}}]{Chaplin2013b}
{Chaplin}, W.~J., {Sanchis-Ojeda}, R., {Campante}, T.~L., {et~al.} 2013, \apj,
  766, 101

\bibitem[{{Cheng} {et~al.}(2012){Cheng}, {Rockosi}, {Morrison},
  {Sch{\"o}nrich}, {Lee}, {Beers}, {Bizyaev}, {Pan}, \&
  {Schneider}}]{Cheng2012}
{Cheng}, J.~Y., {Rockosi}, C.~M., {Morrison}, H.~L., {et~al.} 2012, \apj, 746,
  149

\bibitem[{{Chiappini}(2009)}]{Chiappini2009}
{Chiappini}, C. 2009, in IAU Symposium, Vol. 254, The Galaxy Disk in
  Cosmological Context, ed. J.~{Andersen}, {Nordstr{\"o}ara}, B.~{m}, \&
  J.~{Bland-Hawthorn}, 191--196

\bibitem[{{Chiappini}(2015)}]{Chiappini2015b}
{Chiappini}, C. 2015, in EAS Publications Series, Vol.~73, EAS Publications
  Series, 309--341

\bibitem[{{Chiappini} {et~al.}(2015){Chiappini}, {Anders}, {Rodrigues},
  {Miglio}, {Montalb{\'a}n}, {Mosser}, {Girardi}, {Valentini}, {Noels},
  {Morel}, {Minchev}, {Steinmetz}, {Santiago}, {Schultheis}, {Martig}, {da
  Costa}, {Maia}, {Allende Prieto}, {de Assis Peralta}, {Hekker},
  {Theme{\ss}l}, {Kallinger}, {Garc{\'{\i}}a}, {Mathur}, {Baudin}, {Beers},
  {Cunha}, {Harding}, {Holtzman}, {Majewski}, {M{\'e}sz{\'a}ros}, {Nidever},
  {Pan}, {Schiavon}, {Shetrone}, {Schneider}, \& {Stassun}}]{Chiappini2015}
{Chiappini}, C., {Anders}, F., {Rodrigues}, T.~S., {et~al.} 2015, \aap, 576,
  L12

\bibitem[{{Chiappini} {et~al.}(1997){Chiappini}, {Matteucci}, \&
  {Gratton}}]{Chiappini1997}
{Chiappini}, C., {Matteucci}, F., \& {Gratton}, R. 1997, \apj, 477, 765

\bibitem[{{Christensen-Dalsgaard}(2016)}]{Christensen-Dalsgaard2016}
{Christensen-Dalsgaard}, J. 2016, ArXiv e-prints

\bibitem[{{Cirasuolo} {et~al.}(2014){Cirasuolo}, {Afonso}, {Carollo}, {Flores},
  {Maiolino}, {Oliva}, {Paltani}, {Vanzi}, {Evans}, {Abreu}, {Atkinson},
  {Babusiaux}, {Beard}, {Bauer}, {Bellazzini}, {Bender}, {Best}, {Bezawada},
  {Bonifacio}, {Bragaglia}, {Bryson}, {Busher}, {Cabral}, {Caputi}, {Centrone},
  {Chemla}, {Cimatti}, {Cioni}, {Clementini}, {Coelho}, {Crnojevic}, {Daddi},
  {Dunlop}, {Eales}, {Feltzing}, {Ferguson}, {Fisher}, {Fontana}, {Fynbo},
  {Garilli}, {Gilmore}, {Glauser}, {Guinouard}, {Hammer}, {Hastings}, {Hess},
  {Ivison}, {Jagourel}, {Jarvis}, {Kaper}, {Kauffman}, {Kitching}, {Lawrence},
  {Lee}, {Lemasle}, {Licausi}, {Lilly}, {Lorenzetti}, {Lunney}, {Maiolino},
  {Mannucci}, {McLure}, {Minniti}, {Montgomery}, {Muschielok}, {Nandra},
  {Navarro}, {Norberg}, {Oliver}, {Origlia}, {Padilla}, {Peacock}, {Pedichini},
  {Peng}, {Pentericci}, {Pragt}, {Puech}, {Randich}, {Rees}, {Renzini}, {Ryde},
  {Rodrigues}, {Roseboom}, {Royer}, {Saglia}, {Sanchez}, {Schiavon},
  {Schnetler}, {Sobral}, {Speziali}, {Sun}, {Stuik}, {Taylor}, {Taylor},
  {Todd}, {Tolstoy}, {Torres}, {Tosi}, {Vanzella}, {Venema}, {Vitali},
  {Wegner}, {Wells}, {Wild}, {Wright}, {Zamorani}, \&
  {Zoccali}}]{Cirasuolo2014}
{Cirasuolo}, M., {Afonso}, J., {Carollo}, M., {et~al.} 2014, in \procspie, Vol.
  9147, Ground-based and Airborne Instrumentation for Astronomy V, 91470N

\bibitem[{{CoRoT Team}(2016)}]{Corot2016}
{CoRoT Team}. 2016, {Foreword}, ed. {CoRot Team} (EDP Sciences)

\bibitem[{{Corsaro} {et~al.}(2017){Corsaro}, {Lee}, {Garc{\'{\i}}a},
  {Hennebelle}, {Mathur}, {Beck}, {Mathis}, {Stello}, \&
  {Bouvier}}]{Corsaro2017}
{Corsaro}, E., {Lee}, Y.-N., {Garc{\'{\i}}a}, R.~A., {et~al.} 2017, Nature
  Astronomy, 1, 0064

\bibitem[{{Cui} {et~al.}(2012){Cui}, {Zhao}, {Chu}, {Li}, {Li}, {Zhang}, {Su},
  {Yao}, {Wang}, {Xing}, {Li}, {Zhu}, {Wang}, {Gu}, {Luo}, {Xu}, {Zhang},
  {Liu}, {Zhang}, {Yang}, {Cao}, {Chen}, {Chen}, {Chen}, {Chen}, {Chu}, {Feng},
  {Gong}, {Hou}, {Hu}, {Hu}, {Hu}, {Jia}, {Jiang}, {Jiang}, {Jiang}, {Jin},
  {Li}, {Li}, {Li}, {Liu}, {Liu}, {Lu}, {Mao}, {Men}, {Qi}, {Qi}, {Shi},
  {Tang}, {Tao}, {Wang}, {Wang}, {Wang}, {Wang}, {Wang}, {Wang}, {Wang},
  {Wang}, {Wang}, {Wang}, {Wang}, {Wang}, {Xu}, {Xu}, {Yang}, {Yu}, {Yuan},
  {Yuan}, {Zhai}, {Zhang}, {Zhang}, {Zhang}, {Zhao}, {Zhou}, {Zhou}, {Zhu}, \&
  {Zou}}]{Cui2012}
{Cui}, X.-Q., {Zhao}, Y.-H., {Chu}, Y.-Q., {et~al.} 2012, Research in Astronomy
  and Astrophysics, 12, 1197

\bibitem[{{Dalton} {et~al.}(2014){Dalton}, {Trager}, {Abrams}, {Bonifacio},
  {L{\'o}pez Aguerri}, {Middleton}, {Benn}, {Dee}, {Say{\`e}de}, {Lewis},
  {Pragt}, {Pico}, {Walton}, {Rey}, {Allende Prieto}, {Pe{\~n}ate}, {Lhome},
  {Ag{\'o}cs}, {Alonso}, {Terrett}, {Brock}, {Gilbert}, {Ridings}, {Guinouard},
  {Verheijen}, {Tosh}, {Rogers}, {Steele}, {Stuik}, {Tromp}, {Jasko}, {Kragt},
  {Lesman}, {Mottram}, {Bates}, {Gribbin}, {Rodriguez}, {Delgado}, {Martin},
  {Cano}, {Navarro}, {Irwin}, {Lewis}, {Gonzalez Solares}, {O'Mahony},
  {Bianco}, {Zurita}, {ter Horst}, {Molinari}, {Lodi}, {Guerra}, {Vallenari},
  \& {Baruffolo}}]{Dalton2014}
{Dalton}, G., {Trager}, S., {Abrams}, D.~C., {et~al.} 2014, in \procspie, Vol.
  9147, Ground-based and Airborne Instrumentation for Astronomy V, 91470L

\bibitem[{{Davies} \& {Miglio}(2016)}]{Davies2016}
{Davies}, G.~R. \& {Miglio}, A. 2016, Astronomische Nachrichten, 337, 774

\bibitem[{{Davies} {et~al.}(2016){Davies}, {Silva Aguirre}, {Bedding},
  {Handberg}, {Lund}, {Chaplin}, {Huber}, {White}, {Benomar}, {Hekker}, {Basu},
  {Campante}, {Christensen-Dalsgaard}, {Elsworth}, {Karoff}, {Kjeldsen},
  {Lundkvist}, {Metcalfe}, \& {Stello}}]{Davies2016b}
{Davies}, G.~R., {Silva Aguirre}, V., {Bedding}, T.~R., {et~al.} 2016, \mnras,
  456, 2183

\bibitem[{{de~Bruijne} {et~al.}(2005){de~Bruijne}, Perryman, Lindegren,
  {et~al.}}]{LL:GAIA-JDB-022}
{de~Bruijne}, J., Perryman, M., Lindegren, L., {et~al.} 2005, gAIA-JDB-022

\bibitem[{{de Jong} {et~al.}(2014){de Jong}, {Barden}, {Bellido-Tirado},
  {Brynnel}, {Chiappini}, {Depagne}, {Haynes}, {Johl}, {Phillips}, {Schnurr},
  {Schwope}, {Walcher}, {Bauer}, {Cescutti}, {Cioni}, {Dionies}, {Enke},
  {Haynes}, {Kelz}, {Kitaura}, {Lamer}, {Minchev}, {M{\"u}ller}, {Nuza},
  {Olaya}, {Piffl}, {Popow}, {Saviauk}, {Steinmetz}, {Ural}, {Valentini},
  {Winkler}, {Wisotzki}, {Ansorge}, {Banerji}, {Gonzalez Solares}, {Irwin},
  {Kennicutt}, {King}, {McMahon}, {Koposov}, {Parry}, {Sun}, {Walton},
  {Finger}, {Iwert}, {Krumpe}, {Lizon}, {Mainieri}, {Amans}, {Bonifacio},
  {Cohen}, {Fran{\c c}ois}, {Jagourel}, {Mignot}, {Royer}, {Sartoretti},
  {Bender}, {Hess}, {Lang-Bardl}, {Muschielok}, {Schlichter}, {B{\"o}hringer},
  {Boller}, {Bongiorno}, {Brusa}, {Dwelly}, {Merloni}, {Nandra}, {Salvato},
  {Pragt}, {Navarro}, {Gerlofsma}, {Roelfsema}, {Dalton}, {Middleton}, {Tosh},
  {Boeche}, {Caffau}, {Christlieb}, {Grebel}, {Hansen}, {Koch}, {Ludwig},
  {Mandel}, {Quirrenbach}, {Sbordone}, {Seifert}, {Thimm}, {Helmi}, {trager},
  {Bensby}, {Feltzing}, {Ruchti}, {Edvardsson}, {Korn}, {Lind}, {Boland},
  {Colless}, {Frost}, {Gilbert}, {Gillingham}, {Lawrence}, {Legg}, {Saunders},
  {Sheinis}, {Driver}, {Robotham}, {Bacon}, {Caillier}, {Kosmalski}, {Laurent},
  \& {Richard}}]{deJong2014}
{de Jong}, R.~S., {Barden}, S., {Bellido-Tirado}, O., {et~al.} 2014, in
  \procspie, Vol. 9147, Ground-based and Airborne Instrumentation for Astronomy
  V, 91470M

\bibitem[{{De Silva} {et~al.}(2015){De Silva}, {Freeman}, {Bland-Hawthorn},
  {Martell}, {de Boer}, {Asplund}, {Keller}, {Sharma}, {Zucker}, {Zwitter},
  {Anguiano}, {Bacigalupo}, {Bayliss}, {Beavis}, {Bergemann}, {Campbell},
  {Cannon}, {Carollo}, {Casagrande}, {Casey}, {Da Costa}, {D'Orazi}, {Dotter},
  {Duong}, {Heger}, {Ireland}, {Kafle}, {Kos}, {Lattanzio}, {Lewis}, {Lin},
  {Lind}, {Munari}, {Nataf}, {O'Toole}, {Parker}, {Reid}, {Schlesinger},
  {Sheinis}, {Simpson}, {Stello}, {Ting}, {Traven}, {Watson}, {Wittenmyer},
  {Yong}, \& {{\v Z}erjal}}]{deSilva2015}
{De Silva}, G.~M., {Freeman}, K.~C., {Bland-Hawthorn}, J., {et~al.} 2015,
  \mnras, 449, 2604

\bibitem[{{Deheuvels} {et~al.}(2014){Deheuvels}, {Do{\u g}an}, {Goupil},
  {Appourchaux}, {Benomar}, {Bruntt}, {Campante}, {Casagrande}, {Ceillier},
  {Davies}, {De Cat}, {Fu}, {Garc{\'{\i}}a}, {Lobel}, {Mosser}, {Reese},
  {Regulo}, {Schou}, {Stahn}, {Thygesen}, {Yang}, {Chaplin},
  {Christensen-Dalsgaard}, {Eggenberger}, {Gizon}, {Mathis},
  {Molenda-{\.Z}akowicz}, \& {Pinsonneault}}]{Deheuvels2014}
{Deheuvels}, S., {Do{\u g}an}, G., {Goupil}, M.~J., {et~al.} 2014, \aap, 564,
  A27

\bibitem[{{Deheuvels} {et~al.}(2012){Deheuvels}, {Garc{\'{\i}}a}, {Chaplin},
  {Basu}, {Antia}, {Appourchaux}, {Benomar}, {Davies}, {Elsworth}, {Gizon},
  {Goupil}, {Reese}, {Regulo}, {Schou}, {Stahn}, {Casagrande},
  {Christensen-Dalsgaard}, {Fischer}, {Hekker}, {Kjeldsen}, {Mathur}, {Mosser},
  {Pinsonneault}, {Valenti}, {Christiansen}, {Kinemuchi}, \&
  {Mullally}}]{Deheuvels2012}
{Deheuvels}, S., {Garc{\'{\i}}a}, R.~A., {Chaplin}, W.~J., {et~al.} 2012, \apj,
  756, 19

\bibitem[{{DESI Collaboration} {et~al.}(2016{\natexlab{a}}){DESI
  Collaboration}, {Aghamousa}, {Aguilar}, {Ahlen}, {Alam}, {Allen}, {Allende
  Prieto}, {Annis}, {Bailey}, {Balland}, \& et~al.}]{DESI2016a}
{DESI Collaboration}, {Aghamousa}, A., {Aguilar}, J., {et~al.}
  2016{\natexlab{a}}, ArXiv e-prints

\bibitem[{{DESI Collaboration} {et~al.}(2016{\natexlab{b}}){DESI
  Collaboration}, {Aghamousa}, {Aguilar}, {Ahlen}, {Alam}, {Allen}, {Allende
  Prieto}, {Annis}, {Bailey}, {Balland}, \& et~al.}]{DESI2016b}
{DESI Collaboration}, {Aghamousa}, A., {Aguilar}, J., {et~al.}
  2016{\natexlab{b}}, ArXiv e-prints

\bibitem[{{Di Matteo} {et~al.}(2013){Di Matteo}, {Haywood}, {Combes},
  {Semelin}, \& {Snaith}}]{DiMatteo2013}
{Di Matteo}, P., {Haywood}, M., {Combes}, F., {Semelin}, B., \& {Snaith}, O.~N.
  2013, \aap, 553, A102

\bibitem[{{Di Matteo} {et~al.}(2014){Di Matteo}, {Haywood}, {G{\'o}mez}, {van
  Damme}, {Combes}, {Hall{\'e}}, {Semelin}, {Lehnert}, \&
  {Katz}}]{DiMatteo2014}
{Di Matteo}, P., {Haywood}, M., {G{\'o}mez}, A., {et~al.} 2014, \aap, 567, A122

\bibitem[{{Eggenberger} {et~al.}(2012){Eggenberger}, {Montalb{\'a}n}, \&
  {Miglio}}]{Eggenberger2012}
{Eggenberger}, P., {Montalb{\'a}n}, J., \& {Miglio}, A. 2012, \aap, 544, L4

\bibitem[{{Eisenstein} {et~al.}(2011){Eisenstein}, {Weinberg}, {Agol},
  {Aihara}, {Allende Prieto}, {Anderson}, {Arns}, {Aubourg}, {Bailey},
  {Balbinot}, \& et~al.}]{Eisenstein2011}
{Eisenstein}, D.~J., {Weinberg}, D.~H., {Agol}, E., {et~al.} 2011, \aj, 142, 72

\bibitem[{{Freeman}(2012)}]{Freeman2012}
{Freeman}, K. 2012, Astrophysics and Space Science Proceedings, 26, 137

\bibitem[{{Freeman} \& {Bland-Hawthorn}(2002)}]{Freeman2002}
{Freeman}, K. \& {Bland-Hawthorn}, J. 2002, \araa, 40, 487

\bibitem[{{Fuhrmann}(2011)}]{Fuhrmann2011}
{Fuhrmann}, K. 2011, \mnras, 414, 2893

\bibitem[{{Fuhrmann} {et~al.}(2017){Fuhrmann}, {Chini}, {Kaderhandt}, \&
  {Chen}}]{Fuhrmann2017}
{Fuhrmann}, K., {Chini}, R., {Kaderhandt}, L., \& {Chen}, Z. 2017, \mnras, 464,
  2610

\bibitem[{{Gaia Collaboration} {et~al.}(2016{\natexlab{a}}){Gaia
  Collaboration}, {Brown}, {Vallenari}, {Prusti}, {de Bruijne}, {Mignard},
  {Drimmel}, {Babusiaux}, {Bailer-Jones}, {Bastian}, \&
  et~al.}]{Gaia_Brown2016}
{Gaia Collaboration}, {Brown}, A.~G.~A., {Vallenari}, A., {et~al.}
  2016{\natexlab{a}}, \aap, 595, A2

\bibitem[{{Gaia Collaboration} {et~al.}(2016{\natexlab{b}}){Gaia
  Collaboration}, {Prusti}, {de Bruijne}, {Brown}, {Vallenari}, {Babusiaux},
  {Bailer-Jones}, {Bastian}, {Biermann}, {Evans}, \& et~al.}]{Gaia_Prusti2016}
{Gaia Collaboration}, {Prusti}, T., {de Bruijne}, J.~H.~J., {et~al.}
  2016{\natexlab{b}}, \aap, 595, A1

\bibitem[{{Garc{\'{\i}}a} {et~al.}(2010){Garc{\'{\i}}a}, {Mathur}, {Salabert},
  {Ballot}, {R{\'e}gulo}, {Metcalfe}, \& {Baglin}}]{Garcia2010}
{Garc{\'{\i}}a}, R.~A., {Mathur}, S., {Salabert}, D., {et~al.} 2010, Science,
  329, 1032

\bibitem[{{Gibson} {et~al.}(2009){Gibson}, {Courty},
  {S{\'a}nchez-Bl{\'a}zquez}, {Teyssier}, {House}, {Brook}, \&
  {Kawata}}]{Gibson2009}
{Gibson}, B.~K., {Courty}, S., {S{\'a}nchez-Bl{\'a}zquez}, P., {et~al.} 2009,
  in IAU Symposium, Vol. 254, The Galaxy Disk in Cosmological Context, ed.
  J.~{Andersen}, {Nordstr{\"o}ara}, B.~{m}, \& J.~{Bland-Hawthorn}, 445--452

\bibitem[{{Gilmore} {et~al.}(2012){Gilmore}, {Randich}, {Asplund}, {Binney},
  {Bonifacio}, {Drew}, {Feltzing}, {Ferguson}, {Jeffries}, {Micela}, \&
  et~al.}]{Gilmore2012}
{Gilmore}, G., {Randich}, S., {Asplund}, M., {et~al.} 2012, The Messenger, 147,
  25

\bibitem[{{Girardi} {et~al.}(2012){Girardi}, {Barbieri}, {Groenewegen},
  {Marigo}, {Bressan}, {Rocha-Pinto}, {Santiago}, {Camargo}, \& {da
  Costa}}]{Girardi2012}
{Girardi}, L., {Barbieri}, M., {Groenewegen}, M.~A.~T., {et~al.} 2012,
  Astrophysics and Space Science Proceedings, 26, 165

\bibitem[{{Girardi} {et~al.}(2005){Girardi}, {Groenewegen}, {Hatziminaoglou},
  \& {da Costa}}]{Girardi2005}
{Girardi}, L., {Groenewegen}, M.~A.~T., {Hatziminaoglou}, E., \& {da Costa}, L.
  2005, \aap, 436, 895

\bibitem[{{Gizon} {et~al.}(2013){Gizon}, {Ballot}, {Michel}, {Stahn},
  {Vauclair}, {Bruntt}, {Quirion}, {Benomar}, {Vauclair}, {Appourchaux},
  {Auvergne}, {Baglin}, {Barban}, {Baudin}, {Bazot}, {Campante}, {Catala},
  {Chaplin}, {Creevey}, {Deheuvels}, {Dolez}, {Elsworth}, {Garcia}, {Gaulme},
  {Mathis}, {Mathur}, {Mosser}, {Regulo}, {Roxburgh}, {Salabert}, {Samadi},
  {Sato}, {Verner}, {Hanasoge}, \& {Sreenivasan}}]{Gizon2013}
{Gizon}, L., {Ballot}, J., {Michel}, E., {et~al.} 2013, PNAS, 110, 13267

\bibitem[{{G{\'o}rski} {et~al.}(2005){G{\'o}rski}, {Hivon}, {Banday},
  {Wandelt}, {Hansen}, {Reinecke}, \& {Bartelmann}}]{healpix}
{G{\'o}rski}, K.~M., {Hivon}, E., {Banday}, A.~J., {et~al.} 2005, \apj, 622,
  759

\bibitem[{{Grand} {et~al.}(2016){Grand}, {Springel}, {Kawata}, {Minchev},
  {S{\'a}nchez-Bl{\'a}zquez}, {G{\'o}mez}, {Marinacci}, {Pakmor}, \&
  {Campbell}}]{Grand2016}
{Grand}, R.~J.~J., {Springel}, V., {Kawata}, D., {et~al.} 2016, \mnras, 460,
  L94

\bibitem[{{Guedes} {et~al.}(2013){Guedes}, {Mayer}, {Carollo}, \&
  {Madau}}]{Guedes2013}
{Guedes}, J., {Mayer}, L., {Carollo}, M., \& {Madau}, P. 2013, \apj, 772, 36

\bibitem[{{Gustafsson} {et~al.}(2008){Gustafsson}, {Edvardsson}, {Eriksson},
  {J{\o}rgensen}, {Nordlund}, \& {Plez}}]{MARCS}
{Gustafsson}, B., {Edvardsson}, B., {Eriksson}, K., {et~al.} 2008, \aap, 486,
  951

\bibitem[{{Handberg} {et~al.}(2017){Handberg}, {Brogaard}, \&
  {Miglio}}]{Handberg2017}
{Handberg}, R., {Brogaard}, K., \& {Miglio}, A. 2017, ArXiv e-prints

\bibitem[{{Hawkins} {et~al.}(2016){Hawkins}, {Masseron}, {Jofr{\'e}},
  {Gilmore}, {Elsworth}, \& {Hekker}}]{Hawkins2016}
{Hawkins}, K., {Masseron}, T., {Jofr{\'e}}, P., {et~al.} 2016, \aap, 594, A43

\bibitem[{{Hayden} {et~al.}(2015){Hayden}, {Bovy}, {Holtzman}, {Nidever},
  {Bird}, {Weinberg}, {Andrews}, {Majewski}, {Allende Prieto}, {Anders},
  {Beers}, {Bizyaev}, {Chiappini}, {Cunha}, {Frinchaboy},
  {Garc{\'{\i}}a-Her{\'n}andez}, {Garc{\'{\i}}a P{\'e}rez}, {Girardi},
  {Harding}, {Hearty}, {Johnson}, {M{\'e}sz{\'a}ros}, {Minchev}, {O'Connell},
  {Pan}, {Robin}, {Schiavon}, {Schneider}, {Schultheis}, {Shetrone},
  {Skrutskie}, {Steinmetz}, {Smith}, {Wilson}, {Zamora}, \&
  {Zasowski}}]{Hayden2015}
{Hayden}, M.~R., {Bovy}, J., {Holtzman}, J.~A., {et~al.} 2015, \apj, 808, 132

\bibitem[{{Hayden} {et~al.}(2014){Hayden}, {Holtzman}, {Bovy}, {Majewski},
  {Johnson}, {Allende Prieto}, {Beers}, {Cunha}, {Frinchaboy}, {Garc{\'{\i}}a
  P{\'e}rez}, {Girardi}, {Hearty}, {Lee}, {Nidever}, {Schiavon}, {Schlesinger},
  {Schneider}, {Schultheis}, {Shetrone}, {Smith}, {Zasowski}, {Bizyaev},
  {Feuillet}, {Hasselquist}, {Kinemuchi}, {Malanushenko}, {Malanushenko},
  {O'Connell}, {Pan}, \& {Stassun}}]{Hayden2014}
{Hayden}, M.~R., {Holtzman}, J.~A., {Bovy}, J., {et~al.} 2014, \aj, 147, 116

\bibitem[{{Haywood} {et~al.}(2013){Haywood}, {Di Matteo}, {Lehnert}, {Katz}, \&
  {G{\'o}mez}}]{Haywood2013}
{Haywood}, M., {Di Matteo}, P., {Lehnert}, M.~D., {Katz}, D., \& {G{\'o}mez},
  A. 2013, \aap, 560, A109

\bibitem[{{Hekker} \& {Christensen-Dalsgaard}(2016)}]{Hekker2016}
{Hekker}, S. \& {Christensen-Dalsgaard}, J. 2016, ArXiv e-prints

\bibitem[{{Hekker} {et~al.}(2012){Hekker}, {Elsworth}, {Mosser}, {Kallinger},
  {Chaplin}, {De Ridder}, {Garc{\'{\i}}a}, {Stello}, {Clarke}, {Hall}, \&
  {Ibrahim}}]{Hekker2012}
{Hekker}, S., {Elsworth}, Y., {Mosser}, B., {et~al.} 2012, \aap, 544, A90

\bibitem[{{Holmberg} {et~al.}(2007){Holmberg}, {Nordstr{\"o}m}, \&
  {Andersen}}]{Holmberg2007}
{Holmberg}, J., {Nordstr{\"o}m}, B., \& {Andersen}, J. 2007, \aap, 475, 519

\bibitem[{{Howell} {et~al.}(2014){Howell}, {Sobeck}, {Haas}, {Still},
  {Barclay}, {Mullally}, {Troeltzsch}, {Aigrain}, {Bryson}, {Caldwell},
  {Chaplin}, {Cochran}, {Huber}, {Marcy}, {Miglio}, {Najita}, {Smith},
  {Twicken}, \& {Fortney}}]{Howell2014}
{Howell}, S.~B., {Sobeck}, C., {Haas}, M., {et~al.} 2014, \pasp, 126, 398

\bibitem[{{Huber} {et~al.}(2010){Huber}, {Bedding}, {Stello}, {Mosser},
  {Mathur}, {Kallinger}, {Hekker}, {Elsworth}, {Buzasi}, {De Ridder},
  {Gilliland}, {Kjeldsen}, {Chaplin}, {Garc{\'{\i}}a}, {Hale}, {Preston},
  {White}, {Borucki}, {Christensen-Dalsgaard}, {Clarke}, {Jenkins}, \&
  {Koch}}]{Huber2010}
{Huber}, D., {Bedding}, T.~R., {Stello}, D., {et~al.} 2010, \apj, 723, 1607

\bibitem[{{Huber} {et~al.}(2013){Huber}, {Carter}, {Barbieri}, {Miglio},
  {Deck}, {Fabrycky}, {Montet}, {Buchhave}, {Chaplin}, {Hekker},
  {Montalb{\'a}n}, {Sanchis-Ojeda}, {Basu}, {Bedding}, {Campante},
  {Christensen-Dalsgaard}, {Elsworth}, {Stello}, {Arentoft}, {Ford},
  {Gilliland}, {Handberg}, {Howard}, {Isaacson}, {Johnson}, {Karoff},
  {Kawaler}, {Kjeldsen}, {Latham}, {Lund}, {Lundkvist}, {Marcy}, {Metcalfe},
  {Silva Aguirre}, \& {Winn}}]{Huber2013}
{Huber}, D., {Carter}, J.~A., {Barbieri}, M., {et~al.} 2013, Science, 342, 331

\bibitem[{{Huber} {et~al.}(2012){Huber}, {Ireland}, {Bedding}, {Brand{\~a}o},
  {Piau}, {Maestro}, {White}, {Bruntt}, {Casagrande}, {Molenda-{\.Z}akowicz},
  {Silva Aguirre}, {Sousa}, {Barclay}, {Burke}, {Chaplin},
  {Christensen-Dalsgaard}, {Cunha}, {De Ridder}, {Farrington}, {Frasca},
  {Garc{\'{\i}}a}, {Gilliland}, {Goldfinger}, {Hekker}, {Kawaler}, {Kjeldsen},
  {McAlister}, {Metcalfe}, {Miglio}, {Monteiro}, {Pinsonneault}, {Schaefer},
  {Stello}, {Stumpe}, {Sturmann}, {Sturmann}, {ten Brummelaar}, {Thompson},
  {Turner}, \& {Uytterhoeven}}]{Huber2012}
{Huber}, D., {Ireland}, M.~J., {Bedding}, T.~R., {et~al.} 2012, \apj, 760, 32

\bibitem[{{Huber} {et~al.}(2017){Huber}, {Zinn}, {Bojsen-Hansen},
  {Pinsonneault}, {Sahlholdt}, {Serenelli}, {Silva Aguirre}, {Stassun},
  {Stello}, {Tayar}, {Bastien}, {Bedding}, {Buchhave}, {Chaplin}, {Davies},
  {Garcia}, {Latham}, {Mathur}, {Mosser}, \& {Sharma}}]{Huber2017}
{Huber}, D., {Zinn}, J., {Bojsen-Hansen}, M., {et~al.} 2017, ArXiv e-prints

\bibitem[{{Jacobson} {et~al.}(2016){Jacobson}, {Friel}, {J{\'{\i}}lkov{\'a}},
  {Magrini}, {Bragaglia}, {Vallenari}, {Tosi}, {Randich}, {Donati},
  {Cantat-Gaudin}, {Sordo}, {Smiljanic}, {Overbeek}, {Carraro}, {Tautvai{\v
  s}ien{\.e}}, {San Roman}, {Villanova}, {Geisler}, {Mu{\~n}oz},
  {Jim{\'e}nez-Esteban}, {Tang}, {Gilmore}, {Alfaro}, {Bensby}, {Flaccomio},
  {Koposov}, {Korn}, {Pancino}, {Recio-Blanco}, {Casey}, {Costado},
  {Franciosini}, {Heiter}, {Hill}, {Hourihane}, {Lardo}, {de Laverny}, {Lewis},
  {Monaco}, {Morbidelli}, {Sacco}, {Sousa}, {Worley}, \&
  {Zaggia}}]{Jacobson2016}
{Jacobson}, H.~R., {Friel}, E.~D., {J{\'{\i}}lkov{\'a}}, L., {et~al.} 2016,
  \aap, 591, A37

\bibitem[{{Jofr{\'e}} {et~al.}(2016){Jofr{\'e}}, {Jorissen}, {Van Eck},
  {Izzard}, {Masseron}, {Hawkins}, {Gilmore}, {Paladini}, {Escorza},
  {Blanco-Cuaresma}, \& {Manick}}]{Jofre2016}
{Jofr{\'e}}, P., {Jorissen}, A., {Van Eck}, S., {et~al.} 2016, \aap, 595, A60

\bibitem[{{Jones} \& {Wyse}(1983)}]{Jones1983}
{Jones}, B.~J.~T. \& {Wyse}, R.~F.~G. 1983, \aap, 120, 165

\bibitem[{{Jordi} {et~al.}(2010){Jordi}, {Gebran}, {Carrasco}, {de Bruijne},
  {Voss}, {Fabricius}, {Knude}, {Vallenari}, {Kohley}, \&
  {Mora}}]{2010A&A...523A..48J}
{Jordi}, C., {Gebran}, M., {Carrasco}, J.~M., {et~al.} 2010, \aap, 523, A48

\bibitem[{{Kawata} \& {Chiappini}(2016)}]{Kawata2016}
{Kawata}, D. \& {Chiappini}, C. 2016, Astronomische Nachrichten, 337, 976

\bibitem[{{Kippenhahn} {et~al.}(2012){Kippenhahn}, {Weigert}, \&
  {Weiss}}]{Kippenhahn2012}
{Kippenhahn}, R., {Weigert}, A., \& {Weiss}, A. 2012, {Stellar Structure and
  Evolution} (Springer-Verlag Berlin Heidelberg)

\bibitem[{{Kjeldsen} \& {Bedding}(1995)}]{Kjeldsen1995}
{Kjeldsen}, H. \& {Bedding}, T.~R. 1995, \aap, 293, 87

\bibitem[{{Kubryk} {et~al.}(2015){Kubryk}, {Prantzos}, \&
  {Athanassoula}}]{Kubryk2015}
{Kubryk}, M., {Prantzos}, N., \& {Athanassoula}, E. 2015, \aap, 580, A126

\bibitem[{{Kunder} {et~al.}(2017){Kunder}, {Kordopatis}, {Steinmetz},
  {Zwitter}, {McMillan}, {Casagrande}, {Enke}, {Wojno}, {Valentini},
  {Chiappini}, {Matijevi{\v c}}, {Siviero}, {de Laverny}, {Recio-Blanco},
  {Bijaoui}, {Wyse}, {Binney}, {Grebel}, {Helmi}, {Jofre}, {Antoja}, {Gilmore},
  {Siebert}, {Famaey}, {Bienaym{\'e}}, {Gibson}, {Freeman}, {Navarro},
  {Munari}, {Seabroke}, {Anguiano}, {{\v Z}erjal}, {Minchev}, {Reid},
  {Bland-Hawthorn}, {Kos}, {Sharma}, {Watson}, {Parker}, {Scholz}, {Burton},
  {Cass}, {Hartley}, {Fiegert}, {Stupar}, {Ritter}, {Hawkins}, {Gerhard},
  {Chaplin}, {Davies}, {Elsworth}, {Lund}, {Miglio}, \& {Mosser}}]{Kunder2017}
{Kunder}, A., {Kordopatis}, G., {Steinmetz}, M., {et~al.} 2017, \aj, 153, 75

\bibitem[{{Lagarde} {et~al.}(2015){Lagarde}, {Miglio}, {Eggenberger}, {Morel},
  {Montalb{\'a}n}, {Mosser}, {Rodrigues}, {Girardi}, {Rainer}, {Poretti},
  {Barban}, {Hekker}, {Kallinger}, {Valentini}, {Carrier}, {Hareter},
  {Mantegazza}, {Elsworth}, {Michel}, \& {Baglin}}]{Lagarde2015}
{Lagarde}, N., {Miglio}, A., {Eggenberger}, P., {et~al.} 2015, \aap, 580, A141

\bibitem[{{Lagarde} {et~al.}(2017){Lagarde}, {Robin}, {Reyl{\'e}}, \&
  {Nasello}}]{Lagarde2017}
{Lagarde}, N., {Robin}, A.~C., {Reyl{\'e}}, C., \& {Nasello}, G. 2017, \aap,
  601, A27

\bibitem[{{Lebreton} \& {Goupil}(2014)}]{Lebreton2014}
{Lebreton}, Y. \& {Goupil}, M.~J. 2014, \aap, 569, A21

\bibitem[{{Lillo-Box} {et~al.}(2014){Lillo-Box}, {Barrado}, {Moya},
  {Montesinos}, {Montalb{\'a}n}, {Bayo}, {Barbieri}, {R{\'e}gulo}, {Mancini},
  {Bouy}, \& {Henning}}]{Lillo-Box2014}
{Lillo-Box}, J., {Barrado}, D., {Moya}, A., {et~al.} 2014, \aap, 562, A109

\bibitem[{{Majewski} {et~al.}(2016){Majewski}, {APOGEE Team}, \& {APOGEE-2
  Team}}]{Majewski2016}
{Majewski}, S.~R., {APOGEE Team}, \& {APOGEE-2 Team}. 2016, Astronomische
  Nachrichten, 337, 863

\bibitem[{{Majewski} {et~al.}(2015){Majewski}, {Schiavon}, {Frinchaboy},
  {Allende Prieto}, {Barkhouser}, {Bizyaev}, {Blank}, {Brunner}, {Burton},
  {Carrera}, {Chojnowski}, {Cunha}, {Epstein}, {Fitzgerald}, {Garcia Perez},
  {Hearty}, {Henderson}, {Holtzman}, {Johnson}, {Lam}, {Lawler}, {Maseman},
  {Meszaros}, {Nelson}, {Coung Nguyen}, {Nidever}, {Pinsonneault}, {Shetrone},
  {Smee}, {Smith}, {Stolberg}, {Skrutskie}, {Walker}, {Wilson}, {Zasowski},
  {Anders}, {Basu}, {Beland}, {Blanton}, {Bovy}, {Brownstein}, {Carlberg},
  {Chaplin}, {Chiappini}, {Eisenstein}, {Elsworth}, {Feuillet}, {Fleming},
  {Galbraith-Frew}, {Garcia}, {Anibal Garcia-Hernandez}, {Gillespie},
  {Girardi}, {Gunn}, {Hasselquist}, {Hayden}, {Hekker}, {Ivans}, {Kinemuchi},
  {Klaene}, {Mahadevan}, {Mathur}, {Mosser}, {Muna}, {Munn}, {Nichol},
  {O'Connell}, {Robin}, {Rocha-Pinto}, {Schultheis}, {Serenelli}, {Shane},
  {Silva Aguirre}, {Sobeck}, {Thompson}, {Troup}, {Weinberg}, \&
  {Zamora}}]{Majewski2015}
{Majewski}, S.~R., {Schiavon}, R.~P., {Frinchaboy}, P.~M., {et~al.} 2015, ArXiv
  e-prints

\bibitem[{{Martell} {et~al.}(2017){Martell}, {Sharma}, {Buder}, {Duong},
  {Schlesinger}, {Simpson}, {Lind}, {Ness}, {Marshall}, {Asplund},
  {Bland-Hawthorn}, {Casey}, {De Silva}, {Freeman}, {Kos}, {Lin}, {Zucker},
  {Zwitter}, {Anguiano}, {Bacigalupo}, {Carollo}, {Casagrande}, {Da Costa},
  {Horner}, {Huber}, {Hyde}, {Kafle}, {Lewis}, {Nataf}, {Navin}, {Stello},
  {Tinney}, {Watson}, \& {Wittenmyer}}]{Martell2017}
{Martell}, S.~L., {Sharma}, S., {Buder}, S., {et~al.} 2017, \mnras, 465, 3203

\bibitem[{{Martig} {et~al.}(2016){Martig}, {Fouesneau}, {Rix}, {Ness},
  {M{\'e}sz{\'a}ros}, {Garc{\'{\i}}a-Hern{\'a}ndez}, {Pinsonneault},
  {Serenelli}, {Silva Aguirre}, \& {Zamora}}]{Martig2016}
{Martig}, M., {Fouesneau}, M., {Rix}, H.-W., {et~al.} 2016, \mnras, 456, 3655

\bibitem[{{Martig} {et~al.}(2015){Martig}, {Rix}, {Silva Aguirre}, {Hekker},
  {Mosser}, {Elsworth}, {Bovy}, {Stello}, {Anders}, {Garc{\'{\i}}a}, {Tayar},
  {Rodrigues}, {Basu}, {Carrera}, {Ceillier}, {Chaplin}, {Chiappini},
  {Frinchaboy}, {Garc{\'{\i}}a-Hern{\'a}ndez}, {Hearty}, {Holtzman}, {Johnson},
  {Majewski}, {Mathur}, {M{\'e}sz{\'a}ros}, {Miglio}, {Nidever}, {Pan},
  {Pinsonneault}, {Schiavon}, {Schneider}, {Serenelli}, {Shetrone}, \&
  {Zamora}}]{Martig2015}
{Martig}, M., {Rix}, H.-W., {Silva Aguirre}, V., {et~al.} 2015, \mnras, 451,
  2230

\bibitem[{{Masseron} \& {Gilmore}(2015)}]{Masseron2015}
{Masseron}, T. \& {Gilmore}, G. 2015, \mnras, 453, 1855

\bibitem[{{Masseron} \& {Hawkins}(2017)}]{Masseron2017}
{Masseron}, T. \& {Hawkins}, K. 2017, \aap, 597, L3

\bibitem[{{Mathur} {et~al.}(2016){Mathur}, {Garc{\'{\i}}a}, {Huber}, {Regulo},
  {Stello}, {Beck}, {Houmani}, \& {Salabert}}]{Mathur2016}
{Mathur}, S., {Garc{\'{\i}}a}, R.~A., {Huber}, D., {et~al.} 2016, \apj, 827, 50

\bibitem[{{Matteucci}(2001)}]{Matteucci2001}
{Matteucci}, F., ed. 2001, Astrophysics and Space Science Library, Vol. 253,
  {The chemical evolution of the Galaxy}

\bibitem[{{Mazumdar} {et~al.}(2012){Mazumdar}, {Michel}, {Antia}, \&
  {Deheuvels}}]{Mazumdar2012}
{Mazumdar}, A., {Michel}, E., {Antia}, H.~M., \& {Deheuvels}, S. 2012, \aap,
  540, A31

\bibitem[{{Michalik} {et~al.}(2014){Michalik}, {Lindegren}, {Hobbs}, \&
  {Lammers}}]{Michalik2014}
{Michalik}, D., {Lindegren}, L., {Hobbs}, D., \& {Lammers}, U. 2014, \aap, 571,
  A85

\bibitem[{{Miglio} {et~al.}(2016){Miglio}, {Chaplin}, {Brogaard}, {Lund},
  {Mosser}, {Davies}, {Handberg}, {Milone}, {Marino}, {Bossini}, {Elsworth},
  {Grundahl}, {Arentoft}, {Bedin}, {Campante}, {Jessen-Hansen}, {Jones},
  {Kuszlewicz}, {Malavolta}, {Nascimbeni}, \& {Sandquist}}]{Miglio2016}
{Miglio}, A., {Chaplin}, W.~J., {Brogaard}, K., {et~al.} 2016, \mnras, 461, 760

\bibitem[{{Miglio} {et~al.}(2013){Miglio}, {Chiappini}, {Morel}, {Barbieri},
  {Chaplin}, {Girardi}, {Montalb{\'a}n}, {Valentini}, {Mosser}, {Baudin},
  {Casagrande}, {Fossati}, {Silva Aguirre}, \& {Baglin}}]{Miglio2013}
{Miglio}, A., {Chiappini}, C., {Morel}, T., {et~al.} 2013, \mnras, 429, 423

\bibitem[{{Miglio} {et~al.}(2010){Miglio}, {Montalb{\'a}n}, {Carrier}, {De
  Ridder}, {Mosser}, {Eggenberger}, {Scuflaire}, {Ventura}, {D'Antona},
  {Noels}, \& {Baglin}}]{Miglio2010}
{Miglio}, A., {Montalb{\'a}n}, J., {Carrier}, F., {et~al.} 2010, \aap, 520, L6

\bibitem[{{Mikolaitis} {et~al.}(2014){Mikolaitis}, {Hill}, {Recio-Blanco}, {de
  Laverny}, {Allende Prieto}, {Kordopatis}, {Tautvai{\v s}iene}, {Romano},
  {Gilmore}, {Randich}, {Feltzing}, {Micela}, {Vallenari}, {Alfaro}, {Bensby},
  {Bragaglia}, {Flaccomio}, {Lanzafame}, {Pancino}, {Smiljanic}, {Bergemann},
  {Carraro}, {Costado}, {Damiani}, {Hourihane}, {Jofr{\'e}}, {Lardo},
  {Magrini}, {Maiorca}, {Morbidelli}, {Sbordone}, {Sousa}, {Worley}, \&
  {Zaggia}}]{Mikolaitis2014}
{Mikolaitis}, {\v S}., {Hill}, V., {Recio-Blanco}, A., {et~al.} 2014, \aap,
  572, A33

\bibitem[{{Minchev} {et~al.}(2013){Minchev}, {Chiappini}, \&
  {Martig}}]{Minchev2013}
{Minchev}, I., {Chiappini}, C., \& {Martig}, M. 2013, \aap, 558, A9

\bibitem[{{Minchev} {et~al.}(2014){Minchev}, {Chiappini}, \&
  {Martig}}]{Minchev2014}
{Minchev}, I., {Chiappini}, C., \& {Martig}, M. 2014, \aap, 572, A92

\bibitem[{{Minchev} \& {Famaey}(2010)}]{Minchev2010}
{Minchev}, I. \& {Famaey}, B. 2010, \apj, 722, 112

\bibitem[{{Minchev} {et~al.}(2015){Minchev}, {Martig}, {Streich},
  {Scannapieco}, {de Jong}, \& {Steinmetz}}]{Minchev2015}
{Minchev}, I., {Martig}, M., {Streich}, D., {et~al.} 2015, \apjl, 804, L9

\bibitem[{{Minchev} {et~al.}(2017){Minchev}, {Steinmetz}, {Chiappini},
  {Martig}, {Anders}, {Matijevic}, \& {de Jong}}]{Minchev2017}
{Minchev}, I., {Steinmetz}, M., {Chiappini}, C., {et~al.} 2017, \apj, 834, 27

\bibitem[{{Miville-Desch{\^e}nes} \& {Lagache}(2005)}]{IRAS2002}
{Miville-Desch{\^e}nes}, M.-A. \& {Lagache}, G. 2005, \apjs, 157, 302

\bibitem[{{Molenda-{\.Z}akowicz} {et~al.}(2014){Molenda-{\.Z}akowicz},
  {Brogaard}, {Niemczura}, {Bergemann}, {Frasca}, {Arentoft}, \&
  {Grundahl}}]{Molenda2014}
{Molenda-{\.Z}akowicz}, J., {Brogaard}, K., {Niemczura}, E., {et~al.} 2014,
  \mnras, 445, 2446

\bibitem[{{Morel}(2015)}]{Morel2015}
{Morel}, T. 2015, in Astrophysics and Space Science Proceedings, Vol.~39,
  Asteroseismology of Stellar Populations in the Milky Way, ed. A.~{Miglio},
  P.~{Eggenberger}, L.~{Girardi}, \& J.~{Montalb{\'a}n}, 73

\bibitem[{{Morel} {et~al.}(2014){Morel}, {Miglio}, {Lagarde}, {Montalb{\'a}n},
  {Rainer}, {Poretti}, {Eggenberger}, {Hekker}, {Kallinger}, {Mosser},
  {Valentini}, {Carrier}, {Hareter}, \& {Mantegazza}}]{Morel2014}
{Morel}, T., {Miglio}, A., {Lagarde}, N., {et~al.} 2014, \aap, 564, A119

\bibitem[{{Mosser}(2017)}]{Mosser2017}
{Mosser}, B. 2017, in preparation

\bibitem[{{Mosser} \& {Appourchaux}(2009)}]{Mosser2009}
{Mosser}, B. \& {Appourchaux}, T. 2009, \aap, 508, 877

\bibitem[{{Mosser} {et~al.}(2012){Mosser}, {Goupil}, {Belkacem}, {Marques},
  {Beck}, {Bloemen}, {De Ridder}, {Barban}, {Deheuvels}, {Elsworth}, {Hekker},
  {Kallinger}, {Ouazzani}, {Pinsonneault}, {Samadi}, {Stello}, {Garc{\'{\i}}a},
  {Klaus}, {Li}, {Mathur}, \& {Morris}}]{Mosser2012}
{Mosser}, B., {Goupil}, M.~J., {Belkacem}, K., {et~al.} 2012, \aap, 548, A10

\bibitem[{{Naab} \& {Ostriker}(2016)}]{Naab2016}
{Naab}, T. \& {Ostriker}, J.~P. 2016, ArXiv e-prints

\bibitem[{{Nataf}(2016)}]{Nataf2016}
{Nataf}, D.~M. 2016, \pasa, 33, e023

\bibitem[{{Ness} {et~al.}(2016){Ness}, {Hogg}, {Rix}, {Martig}, {Pinsonneault},
  \& {Ho}}]{Ness2016}
{Ness}, M., {Hogg}, D.~W., {Rix}, H.-W., {et~al.} 2016, \apj, 823, 114

\bibitem[{{Noels} \& {Bragaglia}(2015)}]{Noels2015}
{Noels}, A. \& {Bragaglia}, A. 2015, in Astrophysics and Space Science
  Proceedings, Vol.~39, Asteroseismology of Stellar Populations in the Milky
  Way, ed. A.~{Miglio}, P.~{Eggenberger}, L.~{Girardi}, \& J.~{Montalb{\'a}n},
  167

\bibitem[{{Noels} {et~al.}(2016){Noels}, {Montalb{\'a}n}, \&
  {Chiappini}}]{Noels2016}
{Noels}, A., {Montalb{\'a}n}, J., \& {Chiappini}, C. 2016, Astronomische
  Nachrichten, 337, 982

\bibitem[{{Noguchi}(1998)}]{Noguchi1998}
{Noguchi}, M. 1998, \nat, 392, 253

\bibitem[{{Pagel}(2009)}]{Pagel2009}
{Pagel}, B.~E.~J. 2009, {Nucleosynthesis and Chemical Evolution of Galaxies}
  (Cambridge University Press)

\bibitem[{{Pancino} {et~al.}(2017){Pancino}, {Lardo}, {Altavilla}, {Marinoni},
  {Ragaini}, {Cocozza}, {Bellazzini}, {Sabbi}, {Zoccali}, {Donati}, {Heiter},
  {Koposov}, {Blomme}, {Morel}, {S{\'{\i}}mon-D{\'{\i}}az}, {Lobel},
  {Soubiran}, {Montalban}, {Valentini}, {Casey}, {Blanco-Cuaresma},
  {Jofr{\'e}}, {Worley}, {Magrini}, {Hourihane}, {Fran{\c c}ois}, {Feltzing},
  {Gilmore}, {Randich}, {Asplund}, {Bonifacio}, {Drew}, {Jeffries}, {Micela},
  {Vallenari}, {Alfaro}, {Allende Prieto}, {Babusiaux}, {Bensby}, {Bragaglia},
  {Flaccomio}, {Hambly}, {Korn}, {Lanzafame}, {Smiljanic}, {Van Eck}, {Walton},
  {Bayo}, {Carraro}, {Costado}, {Damiani}, {Edvardsson}, {Franciosini},
  {Frasca}, {Lewis}, {Monaco}, {Morbidelli}, {Prisinzano}, {Sacco}, {Sbordone},
  {Sousa}, {Zaggia}, \& {Koch}}]{Pancino2017}
{Pancino}, E., {Lardo}, C., {Altavilla}, G., {et~al.} 2017, \aap, 598, A5

\bibitem[{{P{\'e}rez Hern{\'a}ndez} {et~al.}(2016){P{\'e}rez Hern{\'a}ndez},
  {Garc{\'{\i}}a}, {Corsaro}, {Triana}, \& {De Ridder}}]{Perez2016}
{P{\'e}rez Hern{\'a}ndez}, F., {Garc{\'{\i}}a}, R.~A., {Corsaro}, E., {Triana},
  S.~A., \& {De Ridder}, J. 2016, \aap, 591, A99

\bibitem[{{Pinsonneault} {et~al.}(2014){Pinsonneault}, {Elsworth}, {Epstein},
  {Hekker}, {M{\'e}sz{\'a}ros}, {Chaplin}, {Johnson}, {Garc{\'{\i}}a},
  {Holtzman}, {Mathur}, {Garc{\'{\i}}a P{\'e}rez}, {Silva Aguirre}, {Girardi},
  {Basu}, {Shetrone}, {Stello}, {Allende Prieto}, {An}, {Beck}, {Beers},
  {Bizyaev}, {Bloemen}, {Bovy}, {Cunha}, {De Ridder}, {Frinchaboy},
  {Garc{\'{\i}}a-Hern{\'a}ndez}, {Gilliland}, {Harding}, {Hearty}, {Huber},
  {Ivans}, {Kallinger}, {Majewski}, {Metcalfe}, {Miglio}, {Mosser}, {Muna},
  {Nidever}, {Schneider}, {Serenelli}, {Smith}, {Tayar}, {Zamora}, \&
  {Zasowski}}]{Pinsonneault2014}
{Pinsonneault}, M.~H., {Elsworth}, Y., {Epstein}, C., {et~al.} 2014, \apjs,
  215, 19

\bibitem[{{Plez}(2012)}]{turbospectrum}
{Plez}, B. 2012, {Turbospectrum: Code for spectral synthesis}, Astrophysics
  Source Code Library

\bibitem[{{Quillen} \& {Garnett}(2001)}]{Quillen2001}
{Quillen}, A.~C. \& {Garnett}, D.~R. 2001, in Astronomical Society of the
  Pacific Conference Series, Vol. 230, Galaxy Disks and Disk Galaxies, ed.
  J.~G. {Funes} \& E.~M. {Corsini}, 87--88

\bibitem[{{Quillen} {et~al.}(2009){Quillen}, {Minchev}, {Bland-Hawthorn}, \&
  {Haywood}}]{Quillen2009}
{Quillen}, A.~C., {Minchev}, I., {Bland-Hawthorn}, J., \& {Haywood}, M. 2009,
  \mnras, 397, 1599

\bibitem[{{Rauer} {et~al.}(2014){Rauer}, {Catala}, {Aerts}, {Appourchaux},
  {Benz}, {Brandeker}, {Christensen-Dalsgaard}, {Deleuil}, {Gizon}, {Goupil},
  {G{\"u}del}, {Janot-Pacheco}, {Mas-Hesse}, {Pagano}, {Piotto}, {Pollacco},
  {Santos}, {Smith}, {Su{\'a}rez}, {Szab{\'o}}, {Udry}, {Adibekyan}, {Alibert},
  {Almenara}, {Amaro-Seoane}, {Eiff}, {Asplund}, {Antonello}, {Barnes},
  {Baudin}, {Belkacem}, {Bergemann}, {Bihain}, {Birch}, {Bonfils}, {Boisse},
  {Bonomo}, {Borsa}, {Brand{\~a}o}, {Brocato}, {Brun}, {Burleigh}, {Burston},
  {Cabrera}, {Cassisi}, {Chaplin}, {Charpinet}, {Chiappini}, {Church},
  {Csizmadia}, {Cunha}, {Damasso}, {Davies}, {Deeg}, {D{\'{\i}}az}, {Dreizler},
  {Dreyer}, {Eggenberger}, {Ehrenreich}, {Eigm{\"u}ller}, {Erikson}, {Farmer},
  {Feltzing}, {de Oliveira Fialho}, {Figueira}, {Forveille}, {Fridlund},
  {Garc{\'{\i}}a}, {Giommi}, {Giuffrida}, {Godolt}, {Gomes da Silva},
  {Granzer}, {Grenfell}, {Grotsch-Noels}, {G{\"u}nther}, {Haswell}, {Hatzes},
  {H{\'e}brard}, {Hekker}, {Helled}, {Heng}, {Jenkins}, {Johansen},
  {Khodachenko}, {Kislyakova}, {Kley}, {Kolb}, {Krivova}, {Kupka}, {Lammer},
  {Lanza}, {Lebreton}, {Magrin}, {Marcos-Arenal}, {Marrese}, {Marques},
  {Martins}, {Mathis}, {Mathur}, {Messina}, {Miglio}, {Montalban}, {Montalto},
  {Monteiro}, {Moradi}, {Moravveji}, {Mordasini}, {Morel}, {Mortier},
  {Nascimbeni}, {Nelson}, {Nielsen}, {Noack}, {Norton}, {Ofir}, {Oshagh},
  {Ouazzani}, {P{\'a}pics}, {Parro}, {Petit}, {Plez}, {Poretti}, {Quirrenbach},
  {Ragazzoni}, {Raimondo}, {Rainer}, {Reese}, {Redmer}, {Reffert},
  {Rojas-Ayala}, {Roxburgh}, {Salmon}, {Santerne}, {Schneider}, {Schou},
  {Schuh}, {Schunker}, {Silva-Valio}, {Silvotti}, {Skillen}, {Snellen}, {Sohl},
  {Sousa}, {Sozzetti}, {Stello}, {Strassmeier}, {{\v S}vanda}, {Szab{\'o}},
  {Tkachenko}, {Valencia}, {Van Grootel}, {Vauclair}, {Ventura}, {Wagner},
  {Walton}, {Weingrill}, {Werner}, {Wheatley}, \& {Zwintz}}]{Rauer2014}
{Rauer}, H., {Catala}, C., {Aerts}, C., {et~al.} 2014, Experimental Astronomy,
  38, 249

\bibitem[{{Reddy} {et~al.}(2006){Reddy}, {Lambert}, \& {Allende
  Prieto}}]{Reddy2006}
{Reddy}, B.~E., {Lambert}, D.~L., \& {Allende Prieto}, C. 2006, \mnras, 367,
  1329

\bibitem[{{Reese}(2016)}]{Reese2016}
{Reese}, D.~R. 2016, {AIMS: Asteroseismic Inference on a Massive Scale},
  Astrophysics Source Code Library

\bibitem[{{Rendle} {et~al.}(2017){Rendle}, {Miglio}, \& {Reese}}]{Rendle2017}
{Rendle}, B., {Miglio}, A., \& {Reese}, D. 2017, in preparation

\bibitem[{{Ricker} {et~al.}(2015){Ricker}, {Winn}, {Vanderspek}, {Latham},
  {Bakos}, {Bean}, {Berta-Thompson}, {Brown}, {Buchhave}, {Butler}, {Butler},
  {Chaplin}, {Charbonneau}, {Christensen-Dalsgaard}, {Clampin}, {Deming},
  {Doty}, {De Lee}, {Dressing}, {Dunham}, {Endl}, {Fressin}, {Ge}, {Henning},
  {Holman}, {Howard}, {Ida}, {Jenkins}, {Jernigan}, {Johnson}, {Kaltenegger},
  {Kawai}, {Kjeldsen}, {Laughlin}, {Levine}, {Lin}, {Lissauer}, {MacQueen},
  {Marcy}, {McCullough}, {Morton}, {Narita}, {Paegert}, {Palle}, {Pepe},
  {Pepper}, {Quirrenbach}, {Rinehart}, {Sasselov}, {Sato}, {Seager},
  {Sozzetti}, {Stassun}, {Sullivan}, {Szentgyorgyi}, {Torres}, {Udry}, \&
  {Villasenor}}]{Ricker2015}
{Ricker}, G.~R., {Winn}, J.~N., {Vanderspek}, R., {et~al.} 2015, Journal of
  Astronomical Telescopes, Instruments, and Systems, 1, 014003

\bibitem[{{Rix} \& {Bovy}(2013)}]{Rix2013}
{Rix}, H.-W. \& {Bovy}, J. 2013, \aapr, 21, 61

\bibitem[{{Robin} {et~al.}(2014){Robin}, {Reyl{\'e}}, {Fliri}, {Czekaj},
  {Robert}, \& {Martins}}]{Robin2014}
{Robin}, A.~C., {Reyl{\'e}}, C., {Fliri}, J., {et~al.} 2014, \aap, 569, A13

\bibitem[{{Rodrigues} {et~al.}(2017){Rodrigues}, {Bossini}, {Miglio},
  {Girardi}, {Montalb{\'a}n}, {Noels}, {Trabucchi}, {Coelho}, \&
  {Marigo}}]{Rodrigues2017}
{Rodrigues}, T.~S., {Bossini}, D., {Miglio}, A., {et~al.} 2017, \mnras

\bibitem[{{Rodrigues} {et~al.}(2014){Rodrigues}, {Girardi}, {Miglio},
  {Bossini}, {Bovy}, {Epstein}, {Pinsonneault}, {Stello}, {Zasowski}, {Allende
  Prieto}, {Chaplin}, {Hekker}, {Johnson}, {M{\'e}sz{\'a}ros}, {Mosser},
  {Anders}, {Basu}, {Beers}, {Chiappini}, {da Costa}, {Elsworth},
  {Garc{\'{\i}}a}, {Garc{\'{\i}}a P{\'e}rez}, {Hearty}, {Maia}, {Majewski},
  {Mathur}, {Montalb{\'a}n}, {Nidever}, {Santiago}, {Schultheis}, {Serenelli},
  \& {Shetrone}}]{Rodrigues2014}
{Rodrigues}, T.~S., {Girardi}, L., {Miglio}, A., {et~al.} 2014, \mnras, 445,
  2758

\bibitem[{{Rojas-Arriagada} {et~al.}(2016){Rojas-Arriagada}, {Recio-Blanco},
  {de Laverny}, {Schultheis}, {Guiglion}, {Mikolaitis}, {Kordopatis}, {Hill},
  {Gilmore}, {Randich}, {Alfaro}, {Bensby}, {Koposov}, {Costado},
  {Franciosini}, {Hourihane}, {Jofr{\'e}}, {Lardo}, {Lewis}, {Lind}, {Magrini},
  {Monaco}, {Morbidelli}, {Sacco}, {Worley}, {Zaggia}, \&
  {Chiappini}}]{Rojas-Arriagada2016}
{Rojas-Arriagada}, A., {Recio-Blanco}, A., {de Laverny}, P., {et~al.} 2016,
  \aap, 586, A39

\bibitem[{{Salaris} {et~al.}(2015){Salaris}, {Pietrinferni}, {Piersimoni}, \&
  {Cassisi}}]{Salaris2015}
{Salaris}, M., {Pietrinferni}, A., {Piersimoni}, A.~M., \& {Cassisi}, S. 2015,
  \aap, 583, A87

\bibitem[{{Samadi} {et~al.}(2012){Samadi}, {Belkacem}, {Dupret}, {Ludwig},
  {Baudin}, {Caffau}, {Goupil}, \& {Barban}}]{Samadi2012}
{Samadi}, R., {Belkacem}, K., {Dupret}, M.-A., {et~al.} 2012, \aap, 543, A120

\bibitem[{{Sandquist} {et~al.}(2016){Sandquist}, {Jessen-Hansen}, {Shetrone},
  {Brogaard}, {Meibom}, {Leitner}, {Stello}, {Bruntt}, {Antoci}, {Orosz},
  {Grundahl}, \& {Frandsen}}]{Sandquist2016}
{Sandquist}, E.~L., {Jessen-Hansen}, J., {Shetrone}, M.~D., {et~al.} 2016,
  \apj, 831, 11

\bibitem[{{Schlegel} {et~al.}(1998){Schlegel}, {Finkbeiner}, \& {Davis}}]{SFD}
{Schlegel}, D.~J., {Finkbeiner}, D.~P., \& {Davis}, M. 1998, \apj, 500, 525

\bibitem[{{Sellwood} \& {Binney}(2002)}]{Sellwood2002}
{Sellwood}, J.~A. \& {Binney}, J.~J. 2002, \mnras, 336, 785

\bibitem[{{Sharma} {et~al.}(2011){Sharma}, {Bland-Hawthorn}, {Johnston}, \&
  {Binney}}]{Sharma2011}
{Sharma}, S., {Bland-Hawthorn}, J., {Johnston}, K.~V., \& {Binney}, J. 2011,
  \apj, 730, 3

\bibitem[{{Shen} \& {Li}(2016)}]{Shen2016}
{Shen}, J. \& {Li}, Z.-Y. 2016, Galactic Bulges, 418, 233

\bibitem[{{Silva Aguirre} {et~al.}(2012){Silva Aguirre}, {Casagrande}, {Basu},
  {Campante}, {Chaplin}, {Huber}, {Miglio}, {Serenelli}, {Ballot}, {Bedding},
  {Christensen-Dalsgaard}, {Creevey}, {Elsworth}, {Garc{\'{\i}}a}, {Gilliland},
  {Hekker}, {Kjeldsen}, {Mathur}, {Metcalfe}, {Monteiro}, {Mosser},
  {Pinsonneault}, {Stello}, {Weiss}, {Tenenbaum}, {Twicken}, \&
  {Uddin}}]{SilvaAguirre2012}
{Silva Aguirre}, V., {Casagrande}, L., {Basu}, S., {et~al.} 2012, \apj, 757, 99

\bibitem[{{Silva Aguirre} {et~al.}(2015){Silva Aguirre}, {Davies}, {Basu},
  {Christensen-Dalsgaard}, {Creevey}, {Metcalfe}, {Bedding}, {Casagrande},
  {Handberg}, {Lund}, {Nissen}, {Chaplin}, {Huber}, {Serenelli}, {Stello}, {Van
  Eylen}, {Campante}, {Elsworth}, {Gilliland}, {Hekker}, {Karoff}, {Kawaler},
  {Kjeldsen}, \& {Lundkvist}}]{SilvaAguirre2015}
{Silva Aguirre}, V., {Davies}, G.~R., {Basu}, S., {et~al.} 2015, \mnras, 452,
  2127

\bibitem[{{Snaith} {et~al.}(2015){Snaith}, {Haywood}, {Di Matteo}, {Lehnert},
  {Combes}, {Katz}, \& {G{\'o}mez}}]{Snaith2015}
{Snaith}, O., {Haywood}, M., {Di Matteo}, P., {et~al.} 2015, \aap, 578, A87

\bibitem[{{Sommer-Larsen} {et~al.}(2003){Sommer-Larsen}, {G{\"o}tz}, \&
  {Portinari}}]{Sommer-Larsen2003}
{Sommer-Larsen}, J., {G{\"o}tz}, M., \& {Portinari}, L. 2003, \apj, 596, 47

\bibitem[{{Steinmetz} \& {Mueller}(1994)}]{Steinmetz1994}
{Steinmetz}, M. \& {Mueller}, E. 1994, \aap, 281, L97

\bibitem[{{Steinmetz} {et~al.}(2006){Steinmetz}, {Zwitter}, {Siebert},
  {Watson}, {Freeman}, {Munari}, {Campbell}, {Williams}, {Seabroke}, {Wyse},
  {Parker}, {Bienaym{\'e}}, {Roeser}, {Gibson}, {Gilmore}, {Grebel}, {Helmi},
  {Navarro}, {Burton}, {Cass}, {Dawe}, {Fiegert}, {Hartley}, {Russell},
  {Saunders}, {Enke}, {Bailin}, {Binney}, {Bland-Hawthorn}, {Boeche}, {Dehnen},
  {Eisenstein}, {Evans}, {Fiorucci}, {Fulbright}, {Gerhard}, {Jauregi}, {Kelz},
  {Mijovi{\'c}}, {Minchev}, {Parmentier}, {Pe{\~n}arrubia}, {Quillen}, {Read},
  {Ruchti}, {Scholz}, {Siviero}, {Smith}, {Sordo}, {Veltz}, {Vidrih}, {von
  Berlepsch}, {Boyle}, \& {Schilbach}}]{Steinmetz2006}
{Steinmetz}, M., {Zwitter}, T., {Siebert}, A., {et~al.} 2006, \aj, 132, 1645

\bibitem[{{Stello} {et~al.}(2015){Stello}, {Huber}, {Sharma}, {Johnson},
  {Lund}, {Handberg}, {Buzasi}, {Silva Aguirre}, {Chaplin}, {Miglio},
  {Pinsonneault}, {Basu}, {Bedding}, {Bland-Hawthorn}, {Casagrande}, {Davies},
  {Elsworth}, {Garcia}, {Mathur}, {Di Mauro}, {Mosser}, {Schneider},
  {Serenelli}, \& {Valentini}}]{Stello2015}
{Stello}, D., {Huber}, D., {Sharma}, S., {et~al.} 2015, \apjl, 809, L3

\bibitem[{{Stello} {et~al.}(2016){Stello}, {Vanderburg}, {Casagrande},
  {Gilliland}, {Silva Aguirre}, {Sandquist}, {Leiner}, {Mathieu}, \&
  {Soderblom}}]{Stello2016}
{Stello}, D., {Vanderburg}, A., {Casagrande}, L., {et~al.} 2016, \apj, 832, 133

\bibitem[{{Turon} {et~al.}(2008){Turon}, {Primas}, {Binney}, {Chiappini},
  {Drew}, {Helmi}, {Robin}, \& {Ryan}}]{Turon2008}
{Turon}, C., {Primas}, F., {Binney}, J., {et~al.} 2008, The Messenger, 134, 46

\bibitem[{{Valentini} {et~al.}(2017){Valentini}, {Chiappini}, {Davies},
  {Elsworth}, {Mosser}, {Lund}, {Miglio}, {Chaplin}, {Rodrigues}, {Boeche},
  {Steinmetz}, {Matijevi{\v c}}, {Kordopatis}, {Bland-Hawthorn}, {Munari},
  {Bienaym{\'e}}, {Freeman}, {Gibson}, {Gilmore}, {Grebel}, {Helmi}, {Kunder},
  {McMillan}, {Navarro}, {Parker}, {Reid}, {Seabroke}, {Sharma}, {Siviero},
  {Watson}, {Wyse}, {Zwitter}, \& {Mott}}]{Valentini2017}
{Valentini}, M., {Chiappini}, C., {Davies}, G.~R., {et~al.} 2017, \aap, 600,
  A66

\bibitem[{{Valentini} {et~al.}(2016){Valentini}, {Chiappini}, {Miglio},
  {Montalb{\'a}n}, {Rodrigues}, {Mosser}, {Anders}, {the CoRoT RG Group}, \&
  {GES Consortium}}]{Valentini2016}
{Valentini}, M., {Chiappini}, C., {Miglio}, A., {et~al.} 2016, Astronomische
  Nachrichten, 337, 970

\bibitem[{{Vrard} {et~al.}(2015){Vrard}, {Mosser}, {Barban}, {Belkacem},
  {Elsworth}, {Kallinger}, {Hekker}, {Samadi}, \& {Beck}}]{Vrard2015}
{Vrard}, M., {Mosser}, B., {Barban}, C., {et~al.} 2015, \aap, 579, A84

\bibitem[{{Wang} {et~al.}(2016){Wang}, {Wang}, {Wu}, {Zhao}, {Li}, {Luo},
  {Liu}, {Zhang}, {Hou}, {Wang}, \& {Cao}}]{Wang2016}
{Wang}, L., {Wang}, W., {Wu}, Y., {et~al.} 2016, \aj, 152, 6

\bibitem[{{Wegg} {et~al.}(2015){Wegg}, {Gerhard}, \& {Portail}}]{Wegg2015}
{Wegg}, C., {Gerhard}, O., \& {Portail}, M. 2015, \mnras, 450, 4050

\bibitem[{{Wisnioski} {et~al.}(2015){Wisnioski}, {F{\"o}rster Schreiber},
  {Wuyts}, {Wuyts}, {Bandara}, {Wilman}, {Genzel}, {Bender}, {Davies},
  {Fossati}, {Lang}, {Mendel}, {Beifiori}, {Brammer}, {Chan}, {Fabricius},
  {Fudamoto}, {Kulkarni}, {Kurk}, {Lutz}, {Nelson}, {Momcheva}, {Rosario},
  {Saglia}, {Seitz}, {Tacconi}, \& {van Dokkum}}]{Wisnioski2015}
{Wisnioski}, E., {F{\"o}rster Schreiber}, N.~M., {Wuyts}, S., {et~al.} 2015,
  \apj, 799, 209

\bibitem[{{Wuyts} {et~al.}(2016){Wuyts}, {Wisnioski}, {Fossati}, {F{\"o}rster
  Schreiber}, {Genzel}, {Davies}, {Mendel}, {Naab}, {R{\"o}ttgers}, {Wilman},
  {Wuyts}, {Bandara}, {Beifiori}, {Belli}, {Bender}, {Brammer}, {Burkert},
  {Chan}, {Galametz}, {Kulkarni}, {Lang}, {Lutz}, {Momcheva}, {Nelson},
  {Rosario}, {Saglia}, {Seitz}, {Tacconi}, {Tadaki}, {{\"U}bler}, \& {van
  Dokkum}}]{Wuyts2016}
{Wuyts}, E., {Wisnioski}, E., {Fossati}, M., {et~al.} 2016, \apj, 827, 74

\bibitem[{{Yanny} {et~al.}(2009){Yanny}, {Rockosi}, {Newberg}, {Knapp},
  {Adelman-McCarthy}, {Alcorn}, {Allam}, {Allende Prieto}, {An}, {Anderson},
  {Anderson}, {Bailer-Jones}, {Bastian}, {Beers}, {Bell}, {Belokurov},
  {Bizyaev}, {Blythe}, {Bochanski}, {Boroski}, {Brinchmann}, {Brinkmann},
  {Brewington}, {Carey}, {Cudworth}, {Evans}, {Evans}, {Gates}, {G{\"a}nsicke},
  {Gillespie}, {Gilmore}, {Nebot Gomez-Moran}, {Grebel}, {Greenwell}, {Gunn},
  {Jordan}, {Jordan}, {Harding}, {Harris}, {Hendry}, {Holder}, {Ivans},
  {Ivezi{\v c}}, {Jester}, {Johnson}, {Kent}, {Kleinman}, {Kniazev},
  {Krzesinski}, {Kron}, {Kuropatkin}, {Lebedeva}, {Lee}, {French Leger},
  {L{\'e}pine}, {Levine}, {Lin}, {Long}, {Loomis}, {Lupton}, {Malanushenko},
  {Malanushenko}, {Margon}, {Martinez-Delgado}, {McGehee}, {Monet}, {Morrison},
  {Munn}, {Neilsen}, {Nitta}, {Norris}, {Oravetz}, {Owen}, {Padmanabhan},
  {Pan}, {Peterson}, {Pier}, {Platson}, {Re Fiorentin}, {Richards}, {Rix},
  {Schlegel}, {Schneider}, {Schreiber}, {Schwope}, {Sibley}, {Simmons},
  {Snedden}, {Allyn Smith}, {Stark}, {Stauffer}, {Steinmetz}, {Stoughton},
  {SubbaRao}, {Szalay}, {Szkody}, {Thakar}, {Sivarani}, {Tucker}, {Uomoto},
  {Vanden Berk}, {Vidrih}, {Wadadekar}, {Watters}, {Wilhelm}, {Wyse}, {Yarger},
  \& {Zucker}}]{Yanny2009}
{Yanny}, B., {Rockosi}, C., {Newberg}, H.~J., {et~al.} 2009, \aj, 137, 4377

\end{thebibliography}

 \clearpage
 \onecolumn
 \section*{Affiliations}
\vspace{0cm}
\begin{overpic}[width=\textwidth]{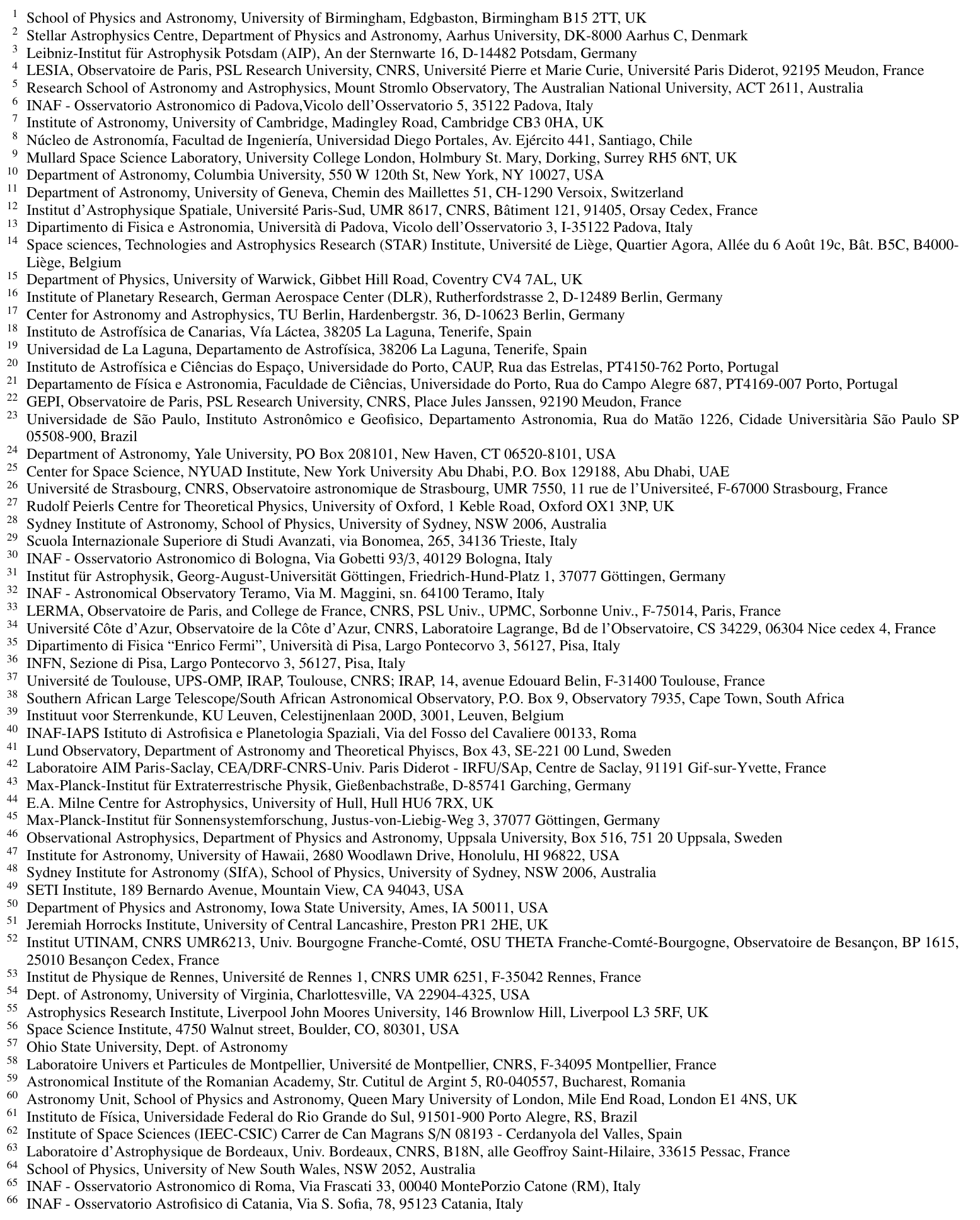}
\end{overpic}
\end{document}